\documentclass[12pt,english]{revtex4-1}
\usepackage[T1]{fontenc}
\usepackage[latin9]{inputenc}
\usepackage{amsmath}
\usepackage{graphicx}
\usepackage{esint}
\usepackage{babel}
\begin{document}
\global\long\def\ket#1{\left|#1\right\rangle }

\global\long\def\bra#1{\left\langle #1\right|}

\global\long\def\braket#1#2{\left\langle #1\left|#2\right.\right\rangle }

\global\long\def\ketbra#1#2{\left|#1\right\rangle \left\langle #2\right|}

\global\long\def\braOket#1#2#3{\left\langle #1\left|#2\right|#3\right\rangle }

\global\long\def\mc#1{\mathcal{#1}}

\global\long\def\nrm#1{\left\Vert #1\right\Vert }

\title{The second law and beyond in microscopic quantum setups}

\author{Raam Uzdin}

\affiliation{Hebrew University of Jerusalem, Jerusalem 9190401, Israel}
\begin{abstract}
The Clausius inequality (CI) is one of the most versatile forms of
the second law. Although it was originally conceived for macroscopic
steam engines, it is also applicable to quantum single particle machines.
Moreover, the CI is the main connecting thread between classical microscopic
thermodynamics and nanoscopic quantum thermodynamics. In this chapter,
we study three different approaches for obtaining the CI. Each approach
shows different aspects of the CI. The goals of this chapter are:
(i) To show the exact assumptions made in various derivations of the
CI. (ii) To elucidate the structure of the second law and its origin.
(iii) To discuss the possibilities each approach offers for finding
additional second-law like inequalities. (iv) To pose challenges related
to the second law in nanoscopic setups. In particular, we introduce
and briefly discuss the notions of exotic heat machines (X machines),
and ``lazy demons''. 
\end{abstract}
\maketitle

\section{Introduction}

Quantum thermodynamics deals with a broad spectrum of issues related
to thermodynamics of small and quantum systems. In particular, this
chapter examines the applicability of one the most fundamental ideas
in thermodynamics: the second law. In addition to its classical role,
the second law is the main connecting thread between classical thermodynamics
and thermodynamics of microscopic (possibly highly quantum) systems.

The second law has various formulations. Here we consider one of the
most versatile and useful forms: the Clausius inequality
\begin{equation}
\Delta S-\intop\frac{\delta Q}{T}\ge0,\label{eq: historic CI}
\end{equation}
where $S$ is the entropy of the system, and $Q$ is the heat exchanged
with a bath at temperature $T$. In classical thermodynamics, entropy
changes are defined by the quantity $\intop\frac{\delta Q}{T}$ evaluated
for a reversible process between two equilibrium states. The bath
is assumed to be large with a well-defined temperature at all stages
of the evolution. 

The Clausius equality (CI) is quite remarkable. It provides a quantitative
prediction that can easily be adapted to the quantum microscopic world.
One of the important results of applying the CI to the microscopic
and/or quantum world, is that the efficiency of quantum heat engines
is limited by the Carnot efficiency even when the evolution exploits
unique quantum properties such as entanglement. The CI also provides
a very easy way to understand Szilard engines, and Landauer's erasure
principle \citep{Landauer1961InfoOrig,reeb2014improved}. 

As described later in detail, under reasonable assumptions the CI
is well understood, and can be obtained for microscopic systems using
several different approaches. However, there are several open questions
and challenges related to the CI: (1) Is the current regime of validity
sufficiently large to handle important microscopic scenarios? If not,
the CI should be further explored and extended. (2) In microscopic
setups where the CI is valid, is it also \emph{useful}? (3) Is the
CI a one of a kind constraint, or is it just one member of a family
of constraints on microscopic thermodynamic processes? Recent studies
show that additional constraints do exist \citep{BrandaoPnasRT2ndLaw,horodecki2013fundamental,LostaglioRudolphCohConstraint,RU2017genCI}. 

The importance of additional constraints or ``laws'' comes from
the fact that in nanoscopic setups, systems are easily taken out of
equilibrium. In equilibrium macroscopic systems, knowledge of a few
coarse-grained properties (e.g., internal energy, volume, pressure,
etc.) is sufficient for knowing almost everything about the system.
In contrast, in small out of equilibrium systems, changes in the internal
energy, for example, provide only a limited amount of information
on how the energy distribution deviates from equilibrium. Moreover,
in small systems, it is possible to experimentally probe features
that are more detailed than the average energy, e.g., the energy variance.
Thus, it is intriguing to ask if there are thermodynamic constraints,
perhaps of the CI form, on other moments of the energy \citep{RU2017genCI},
or on other measurable features of the system. Furthermore, in microscopic
setups, the environment itself may be very small. Thus, it is conceivable
to measure various environment properties (e.g. energy variance) as
the environment starts to deviate from its initial equilibrium state.

In trying to extend the regime of validity of the CI, or in the search
of new thermodynamic constraints on new quantities, it is important
to keep track on what properties of the CI have been retained, what
features have been lost or replaced, and at what cost. In particular,
does the extension involve measurable quantities and what is the predictive
power of the new suggested extension with respect to present and future
experiments. For example, thermodynamic resource theory predicts that
there are families of mathematical constraints on a specific set of
processes involving thermal baths. This is a very interesting and
important finding. However, these constraints do not deal directly
with observable quantities and their operational (or informational)
meaning has not been clarified yet. It is not a necessary requirement
that new constraints will have the same form and nature of the CI.
However, it is very interesting to explore and find constraints that
resemble the second law and its rock-solid, well-established logic.
Thus, this chapter puts emphasis on the structure of the CI, and its
various features.

In what follows, we aim to achieve the goals stated in the abstract
by studying three different approached to derive the CI: (1) The reduced
entropy approach; (2) The global passivity approach; (3) The Completely
positive maps approach. Each approach shows different aspects of the
CI, and can lead to new challenges related to the second law in nanoscopic
setups. Before exploring each approach let us state the CI statement
of the second law in microscopic setups.

\section{The CI in microscopic and quantum setups\label{sec: CI in micro setups}}

\begin{figure}
\includegraphics[width=10cm]{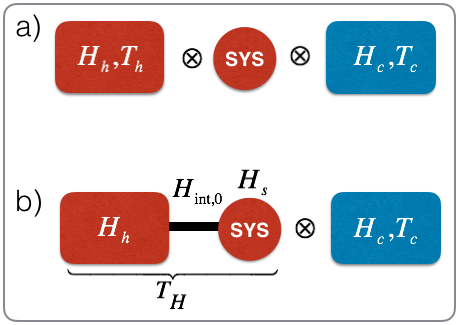}

\caption{\label{fig: init_setups}(a) The standard CI form of the second law
assumes that all elements are initially uncorrelated (\ref{eq: basic r0}).
(b) In microscopic setups where a system is initially strongly coupled
to one of the environments, the system starts in a \textit{coupled}
Gibbs state (\ref{eq: r0 coupled Gibbs}). Hence, in general, the
system does not have a thermal form then the environment is traced
out. The CI does not hold in this scenario. Nonetheless, new approaches
can successfully handle this scenario (Sec. \ref{subsec: CCI}).}
\end{figure}

Consider a setup that is composed of a system that is initially in
an arbitrary mixed or pure state, and several environments that are
initially in thermal equilibrium (Gibbs states). All elements ('element'
may refer to the system, or to one of its environments) are initially
uncorrelated, so at $t_{0}$ the setup shown in Fig. 1a is described
by the following total density matrix (Fig. \ref{fig: init_setups}a):
\begin{equation}
\rho_{0}^{tot}=\rho_{0}^{sys}\otimes\frac{e^{-\beta_{1}H_{1}}}{Z_{1}}\otimes\frac{e^{-\beta H_{2}}}{Z_{2}}\otimes...\label{eq: basic r0}
\end{equation}
where $H_{k}$ is the Hamiltonian of environment $k$, $\beta_{k}=1/T_{k}$
is its \textit{initial} inverse temperature, and the $Z_{i}$'s are
normalization factors. We call an environment that is initially prepared
in a thermal state a 'microbath' (later denoted in equations by $\mu b$).
Unlike macroscopic baths, microbaths are not assumed to be large and
to remain in a thermal state while interacting with the system. The
only requirement of a microbath is that its initial state is thermal.
That said, the microbath is allowed to be very large. The preparation
process of a microbath is outside the scope of the present discussion.
We shall assume that such (possibly small) initially thermal environments
are given in the beginning of the experiment. Nevertheless, in principle,
they can be prepared by weakly interacting with a much larger thermal
environment.

The evolution of the setup is generated by some global time-dependent
Hamiltonian that describes all the interactions between the system,
the microbaths, and external fields (i.e., driving). In such setups
the following microscopic form of the CI \citep{PeresBook,Esposito2011EPL2Law,Sagawa2012second,strasberg2017quantum,lostaglio2017non-commutativity,YelenaGGE,halpernGGErt}
holds:

\begin{align}
\Delta S_{VN}^{sys}+\sum\beta_{k}q_{k} & \ge0\label{eq: basic CI}\\
q_{k} & =\Delta\left\langle H_{k}\right\rangle 
\end{align}
where $S_{VN}^{sys}=-tr[\rho^{sys}\ln\rho^{sys}]$ is the von Neumann
entropy of the system (we shall drop the VN subscript hereafter).
\textit{The $\Delta$ always refers to changes with respect to time
$t_{0}$ where (\ref{eq: basic r0}) holds}, e.g., at time $t$: $\Delta S^{sys}=S(t)-S(t_{0})$.
$q_{k}$ is the heat or more accurately the energy transferred to
$k$-th microbath. For $q_{k}$ to represent the energy change in
bath k, $H_{k}$ must be satisfy $H_{k}(t_{f})=H_{k}(t_{0})$ which
means that at the end of the process the environment is not modified
directly by external field (it only interacts with other elements).
A classical derivation is carried out in \citep{JarzynskiMicroscopicClausius,deffner2013information}.
We point out a few interesting facts on the microscopic CI result
(\ref{eq: basic CI})
\begin{itemize}
\item As long as the initial condition (\ref{eq: basic r0}) holds, the
interaction between the elements and the external driving can be arbitrarily
strong and lead to highly non-Markovian dynamics.
\item The microbaths and the system can be arbitrary small in size (e.g.,
a single spin) as long as the initial state of the setup is given
by (\ref{eq: basic r0}).
\item The microbath can be very far from thermal equilibrium at the end
and during the interaction with the system. The $\beta_{k}$'s refer
only to the \textit{initial} temperatures of the microbath.
\end{itemize}
For additional refinements of the CI see \citep{reeb2014improved,DeffnerLutzRelEntBures}.
In what follows, we derive (\ref{eq: basic CI}) using different approaches.

\section{The reduced entropy approach\label{sec:The-reduced-entropy}}

For simplicity of notation, let us assume we have only three elements:
a system '$s$', and two environments '$A$' and '$B$'. At this point,
it is not assumed that the environments are initially thermal. However,
all three elements are assumed to be initially uncorrelated to each
other so the initial density matrix is given by

\begin{equation}
\rho_{0}^{tot}=\rho_{0}^{s}\otimes\rho_{0}^{A}\otimes\rho_{0}^{B}.\label{eq: rho tot 0}
\end{equation}
Next, we assume that the evolution is generated by some time-dependent
Hamiltonian of the form:
\begin{equation}
H_{tot}(t)=H_{s}(t)+H_{A}(t)+H_{B}(t)+H_{int}(t),\label{eq: Htot}
\end{equation}
where the first term describes possible external driving of the system
(e.g. by laser light), the next two terms are the environment Hamiltonians
(which may also be subjected to external fields), and the last term
describes the time-dependent interactions between the three elements.
This generic Hamiltonian can describe almost any thermodynamic protocol
that does not involve feedback or measurements during the evolution.
No other interactions or other parties are involved. Note that typically
the environments ($A$ and $B$) are either not driven $\frac{d}{dt}H_{A(B)}=0$
by external fields, or at the very least are not modified at the end
of the thermodynamic protocol
\begin{equation}
H_{A(B)}(t_{f})=H_{A(B)}(t_{0}).\label{eq: no env drive}
\end{equation}
However, this assumption is not needed until Eq. (\ref{eq: q def}).

In quantum mechanics, any $H_{tot}(t)$ leads to a unitary evolution
operator $U$ (i.e., $U^{\dagger}U=I$) that relates the final density
matrix to the initial one,

\begin{equation}
\rho_{t}^{tot}=U\rho_{0}^{tot}U^{\dagger}.\label{eq: uni evo}
\end{equation}

Finally, we consider a slightly more general case where there is some
noise in $H_{tot}(t)$ (this noise can be a control amplitude noise,
timing noise, etc) so that with probability $p_{j}$, $U_{j}$ is
executed instead of the desired $U$. As a result, the evolution of
the setup is described by a mixture of unitaries
\begin{equation}
\rho_{t}^{tot}=\sum p_{k}U_{k}\rho_{0}^{tot}U_{k}^{\dagger}.\label{eq: mix of uni}
\end{equation}
Next, we use the non-negativity of the quantum relative entropy \citep{nielsen2002QuantInfo}
for any two density matrices $\rho$ and $\sigma$

\begin{equation}
D(\rho|\sigma)\doteq\text{tr}[\rho(\ln\rho-\ln\sigma)]\ge0,\label{eq: rel ent def}
\end{equation}
where equality holds if and only if $\rho=\sigma$. We set $\rho$
to be the state of the setup at time t $\rho=\rho_{t}^{tot}$, and
$\sigma$ to be the tensor product of the local density matrices $\sigma=\rho_{t}^{s}\otimes\rho_{t}^{A}\otimes\rho_{t}^{B}$
where $\rho_{t}^{(i)}\doteq\text{tr}_{\neq i}[\rho_{t}^{tot}]$. Using
the log property for\textit{ product states }\textit{\emph{($\sigma$)}}\textit{,}
we get

\begin{equation}
D(\rho_{t}^{tot}|\rho_{t}^{s}\otimes\rho_{t}^{A}\otimes\rho_{t}^{B})=(\sum_{j=s,A,B}S_{t}^{j})-S_{t}^{tot}\ge0,\label{eq: D tot prod 2}
\end{equation}
where the inequality follows from the non negativity of the quantum
relative entropy (\ref{eq: rel ent def}). If the global evolution
is unitary (\ref{eq: uni evo}), then $S_{t}^{tot}=S_{0}^{tot}$ since
all the eigenvalues of the total density matrix are conserved quantities.
If the global evolution is a mixture of unitaries (\ref{eq: mix of uni})
then due to the Schur concavity \citep{Marshall1979MajorizationBook}
of the von Neumann entropy it holds that,
\begin{equation}
S_{t}^{tot}\ge S_{0}^{tot}.\label{eq: Growth Stot}
\end{equation}
Note that this inequality is \textit{not} a manifestation of the second
law. Unlike the inequality (\ref{eq: D tot prod 2}) that holds also
for a unitary evolution, the one in (\ref{eq: Growth Stot}) originates
from the randomness/noise in the protocol that we have included for
generality. Using (\ref{eq: Growth Stot}) in (\ref{eq: D tot prod 2})
we get

\begin{equation}
(\sum_{j=s,A,B}S_{t}^{j})-S_{0}^{tot}\ge0.\label{eq:dS stot0}
\end{equation}
Since the setup starts in a product state (\ref{eq: rho tot 0}) it
holds that $S_{0}^{tot}=\sum_{j=s,c,h}S_{0}^{j}$ and (\ref{eq:dS stot0})
becomes the reduced entropy growth form of the (microscopic) second
law \citep{PeresBook,Esposito2011EPL2Law,Sagawa2012second},
\begin{equation}
\sum_{j=s,c,h}\Delta S^{j}\ge0.\label{eq: DS^j}
\end{equation}
This form is strictly entropic and has no reference to heat or energy.
Moreover, (\ref{eq: DS^j}) holds even if the environments are initially
in highly non-thermal states. For two parties, (\ref{eq: DS^j}) can
be obtained from the non-negativity of the quantum mutual information
\citep{nielsen2002QuantInfo}. When there are more parties, (\ref{eq: D tot prod 2})
should be used.

Equation (\ref{eq: DS^j}) has the following information interpretation.
For any global unitary the total entropy is conserved $\Delta S^{tot}=0$.
However, the final density matrix contains classical and quantum correlations
between the different elements. The sum of reduced entropies does
not include these correlations, and therefore it grows despite the
fact that the total entropy is conserved. 

It is important to highlight several points. (1) This growth is only
with respect to the initial values, and the growth is typically non-monotonic
in time due to the possible non-Markovian nature of the evolution.
(2) The reason the entropy is growing and not decreasing is due to
the special assumption that the elements are initially uncorrelated.
(3) As mentioned earlier, no assumptions were made on the size of
the elements. The bath or the system can be as small as a single spin.
These features make the entropic form of the second law useful for
understanding processes involving small quantum systems (e.g.,  algorithmic
cooling \citep{tal02}).

\subsection[The energy-information form]{The Clausius inequality: the energy-information form of the second
law\label{subsec:Energy-information-form}}

The transition from the entropic form of the second law to the energy-information
CI form, involves three more simple steps. The first is to apply the
following identity that holds for any two density matrices $\rho_{1},\rho_{2}$.
\begin{equation}
S(\rho_{2})-S(\rho_{1})\equiv\text{tr}[(\rho_{2}-\rho_{1})(-\ln\rho_{1})]-D(\rho_{2}|\rho_{1}).\label{eq: basic DS identity}
\end{equation}
Applying this identity to the initial and final reduced states of
the $k$-th environment
\begin{equation}
\Delta S^{env,k}=\Delta\left\langle -\ln\rho_{0}^{env,k}\right\rangle -D(\rho_{f}^{k}|\rho_{0}^{k}),
\end{equation}
and substituting it in (\ref{eq: DS^j}) we obtain
\begin{equation}
\Delta S^{sys}+\sum_{k}\Delta\left\langle -\ln\rho_{0}^{env,k}\right\rangle \ge\sum_{k}D(\rho_{f}^{k}|\rho_{0}^{k}).\label{eq: dS B D}
\end{equation}
The next step is to set the environments to be microbaths, i.e. $\rho_{0}^{env,k}\to\rho_{0}^{\mu b,k}=e^{-\beta_{k}H_{k}^{\mu b}}/Z_{k}$
and get
\begin{equation}
\Delta S^{sys}+\sum_{k}\beta_{k}q_{k}\ge\sum_{k}D(\rho_{f}^{\mu b,k}|\rho_{0}^{\mu b,k}),\label{eq: CI with Dbath}
\end{equation}
where, as before, we denote the change in the average energy of bath
$k$ by $q_{k}$
\begin{equation}
q_{k}\doteq\text{tr}[(\rho_{f}^{\mu b,k}-\rho_{0}^{\mu b,k})H_{k}^{\mu b}].\label{eq: q def}
\end{equation}
In interpreting $q_{k}$ as the change in the energy of the bath we
use assumption (\ref{eq: no env drive}). Due to the the relative
entropy terms in (\ref{eq: CI with Dbath}), form (\ref{eq: CI with Dbath})
is slightly stronger than the standard CI (\ref{eq: basic CI}). However,
the relative entropy terms cannot be easily measured. Nevertheless,
they are not void of physical meaning. They represent the work that
can be extracted from the microbaths due to the fact they are not
in a thermal state at the end of the process. If, for example, \textit{an
additional auxiliary large bath} at temperature $T_{h}$ was available,
then a $T_{h}D(\rho_{f}^{\mu b,h}|\rho_{0}^{\mu b,h})$ amount of
work could have been reversibly extracted from the hot microbath.
However, since we consider an entropically self-contained setups (where
a large external bath is not included), this work extraction scheme
is irrelevant. 

In the last third step, the relative entropy property $D(\rho_{f}^{k}|\rho_{0}^{k})\ge0$
is used to obtain the CI inequality (\ref{eq: basic CI}) from (\ref{eq: CI with Dbath}).
Finally, by defining the operator
\begin{align}
B^{env} & =-\ln\rho_{0}^{env},
\end{align}
where $\rho_{0}^{baths}$ is the initial density matrices of all the
microbaths in the setup, the CI can be written more compactly as
\begin{equation}
\Delta S^{sys}+\Delta\left\langle B^{env}\right\rangle \ge0.\label{eq: CI Bbath}
\end{equation}
This form will be used in Sec. \ref{subsec: CCI}.

\subsection{The structure of the CI\label{subsec: CI-Structure}}

The CI connects two very different quantities. One is an information
measure, and the other is an expectation value obtained from energy
measurements. These two quantities have completely different nature
and properties. $S^{sys}(\rho^{sys})$ is an information measure and
as such it is nonlinear in $\rho_{sys}$. It is basis-independent,
and invariant to permutations and unitary transformation: $S^{sys}(\rho^{sys})=S^{sys}(U\rho^{sys}U^{\dagger})$.
That is, only the eigenvalues of the density matrix are important.
It matters not to which orthogonal states these eigenvalues are assigned.
$S^{sys}(\rho^{sys})$ quantifies the ignorance about the system.
If it is in a known pure state $\ket{\psi}$ then $S^{sys}(\ketbra{\psi}{\psi})=0$.
If $\rho^{sys}$ is in a fully mixed state then $S^{sys}$ obtains
a maximal value. 

The second term in the CI $\beta q=\beta\Delta\left\langle H_{\mu b}\right\rangle $
is completely different. First, it is linear in the density matrix.
Second, it is not invariant under permutations and more general unitary
transformations $tr[U\rho^{\mu b}U^{\dagger}H^{\mu b}]\neq tr[\rho^{\mu b}H^{\mu b}]$.
Moreover, $\left\langle H^{\mu b}\right\rangle $ involves only the
diagonal elements of the density matrix in the energy basis. Thus,
dephasing all the coherences in the energy basis will not change the
energy expectation values at a given instant (it will, however, change
the evolution from that point on). In contrast, information quantities
such as the von Neumann entropy are very sensitive to dephasing. All
types of dephasing operations increase the values of $S^{sys}(\rho^{sys})$.
We point out that all Schur concave information measures have this
property \citep{Marshall1979MajorizationBook,RU2017genCI}.

This relation between a quantity that is basis-independent and a quantity
that is basis-dependent can be quite powerful. For example, if we
erase one bit of information from our system (no matter in which basis),
then according to the CI (Landauer principle \citep{Landauer1961InfoOrig,reeb2014improved})
at least $T\ln2$ amount of heat has to be exchanged with the bath.
Thus, although nonlinear quantities such as $S^{sys}(\rho^{sys})$
are not directly measurable (they are not associated with an Hermitian
operator), they can be quite insightful when associated with expectation
values as in the CI. Note that in the CI the nonlinear quantities
involve only the system, which at least in nanoscopic setups can be
regarded as small. For example, it is not unreasonable to perform
tomography of a two-spin system, if needed. On the other hand, it
not practical to perform a full tomography of a thirty-spin microbath
in order to evaluate its entropy changes. Hence, when using nonlinear
quantities in thermodynamics, we should be mindful how they can be
evaluated or used either directly or indirectly.

Finally, we point out that a major feature of the CI is that the inequality
is saturated (becomes an equality) for reversible processes. This
saturation is highly important for two main reasons. First, inequalities
that are not saturated can be completely useless. For example, $\Delta S+\beta q+10^{100}\ge0$
is also a valid inequality, but it is trivially satisfied and cannot
be used to state something useful on $\Delta S$ by knowing $q$ or
vice versa. The second reason why the CI saturation is important is
that it provides a special meaning to reversible processes compared
to irreversible processes. All reversible processes between two endpoints
(two density matrices of the systems) are equivalent in terms of heat,
work, and entropy changes in the bath. On the other hand, irreversible
processes are path-dependent and lead to suboptimal work extraction.
As a side note, we point out that creating a reversible interaction
with microbaths is not as straightforward as it is in macroscopic
setups since, in general, the microbaths do not remain in a Gibbs
state. 

\subsection[Fluctuation theorems and the CI]{Fluctuation theorems and Clausius-like inequalities\label{subsec: FT vs CI}}

Fluctuation theorems (FT's) are intensively studied for the last few
decades (see reviews \citep{Jarzynski2011equalitiesReview,Seifert2007FTRev,harris2007fluctuationReview}).
Their main appeal is that they provide \textit{equalities} that hold
even when the system is driven far away from equilibrium. In contrast,
the CI provides a potentially saturated \emph{inequality} in non-equilibrium
dynamics. 

Interestingly, by applying mathematical inequality (Jensen inequality),
some manifestations of the CI can be retrieved from FT's. For this
reason, it is sometimes claimed that FT's are more fundamental than
the CI. However, it is highly important to remember that, presently,
the CI is applicable in various scenarios where FT's may not hold
or become impractically to use: (1) Initial coherence in the energy
eigenbasis of the systems. (2) Interaction with microbaths that substantially
deviate from thermal equilibrium during the evolution so that detailed
balance does not hold anymore. In addition, FT's yields the a weaker
form of the second law which is based on the equilibrium entropy (equilibrium
free energy) and not in terms of von Neumann entropy of the final
non-equilibrium state (non-equilibrium free energy (\ref{eq: W DF CI})).
Moreover, treating multiple bath in FT's, as in the CI, is also a
non trivial matter. In particular, the observed quantity is no longer
work \citep{campisi2014FT_SolidStateExp}. 

In summary, in comparison to the CI, fluctuation theorems presently
provide stronger statements (equalities) but for more limited physical
scenarios. Further study is needed to understand if there is a fundamental
complementarity between the strength of non-equilibrium results (FT
or CI) and their regime of validity, or perhaps it is possible to
hold the stick at both ends and find FT's that are both more general
and stronger than the CI.

Finally, we point that energy appears in the CI in the form of ``difference
of averages'' $q=\left\langle H_{\mu b}\right\rangle _{f}-\left\langle H_{\mu b}\right\rangle _{0}$,
while in FT (e.g. \citep{jarzynski1997nonequilibrium,campisi2014FT_SolidStateExp})
energy appears as `` nonlinear average of differences'', e.g., $\left\langle e^{-\beta(E_{f}-E_{0})}\right\rangle $
where $E_{f}$ and $E_{0}$ are initial and final energy measurements
in a specific realization (trajectory) \citep{jarzynski1997nonequilibrium}.
That is, the energy term in the CI has the structure $\Delta\left\langle g(H)\right\rangle $
while in FT's the structure is $\left\langle \tilde{g}(\Delta E)\right\rangle $
where $g$ and $\tilde{g}$ are some analytic functions. In FT's the
evaluation of $\Delta E$ involves measuring $E_{0}$ and $E_{f}$
at the same run of the experiment. In quantum mechanics $E_{0}$ measurements
will modify the evolution due to the loss of coherence (wavefunction
``collapse''). In contrast, quantities such $\left\langle g(H)\right\rangle _{f}$
and $\left\langle g(H)\right\rangle _{0}$ are evaluated in different
runs of the experiment: in one set experimental runs $\left\langle g(H)\right\rangle _{0}$
is measured and evaluated, and in another set of runs $\left\langle g(H)\right\rangle _{f}$
is measured. Therefore, the structure of $\Delta\left\langle g(H)\right\rangle $
is more compatible with quantum mechanics compared to $\left\langle \tilde{g}(\Delta E)\right\rangle $
that appears in FT's. 

\subsection[CI Deficiencies]{\label{subsec: Two-deficiencies}Deficiencies and limitations of
the CI}

Despite its great success and its internal consistency, the CI also
has a few deficiencies. One deficiency concerns very cold environments
and the other occurs in dephasing interactions. These deficiencies
are not inconsistencies but scenarios where the CI provides trivial
and useless predictions. 

In the limit of a very low temperature, the predictive power of the
Clausius inequality is degraded. As $T=\frac{1}{\beta}\to0$, the
$\beta q$ term diverges and the changes in the entropy of the system
becomes negligible (the entropy changes in the system are bounded
by the logarithm of the system's Hilbert dimension). Hence, in this
limit, the CI reads $\beta q\ge0$, which means that heat is flowing
into this initially very cold bath. This is hardly a surprise. If
the initial temperature is very low the microbath will be in the ground
state. Thus, it is clear that any interaction with other agents can
only increase its average energy.

Another trivial, and somewhat useless prediction of the CI is obtained
in dephasing interactions. Such interactions satisfy $[H_{int},H_{sys}]=0$
and typically also $[H_{int},H_{env}]=0$. Consequently, dephasing
interactions do not affect the energy distributions of the system
and the environment. However, they degrade the coherence in the energy
basis of the system. Since there is no heat flow involved, the CI
predicts $\Delta S^{sys}\ge0$. This is a trivial result: any dephasing
process can be written as a mixture of unitaries operating on the
system only, and mixture of unitaries always increases the entropy
due to the concavity of the von Neumann entropy. 

In addition to these two deficiencies, two more obvious deficiencies
should be mentioned. The first is the restriction to initially uncorrelated
system and environment. In some important microscopic scenarios, this
assumption is not valid. This is not a limitation of the derivation
in \ref{subsec:Energy-information-form} since in the presence of
an initial correlation it is easy to find scenarios which indeed violates
the CI (e.g., \citep{SerraLutz2spinNMR}). Another limitation of the
CI is that in some cases it does not deal with the quantities of interest.
For example in Sec. \ref{subsec: X-machines} we discuss machines
whose output cannot be expressed in terms of average energy changes
or entropy changes. Hence, the CI does not impose a performance limit
for these machines.

\subsection[The passive CI]{The passive CI form: environment energy vs. heat}

In this chapter we refer to $q$, the energy exchanged with the bath,
as heat. However, this terminology ignores the fact that the microbath
can be in a non-passive state that admits work extraction by applying
a local unitary on the microbath. Moreover, some studies suggested
using 'squeezed thermal bath' as fuel for quantum heat machines \citep{LutzSqueezedBaths,ParrondoSqueezedBath,Roulet_3_Ion_fridge2017}.
A squeezed microbath is obtained by applying a unitary $U_{\mu b}$
to a microbath $\rho_{\beta,sq}^{\mu b}=U_{\mu b}\rho_{\beta}^{\mu b}U_{\mu b}^{\dagger}$.
Since the thermal state has the lowest average energy for a given
von Neumann entropy, any squeezing operation of a thermal state increases
the energy of the bath. Thus, the preparation of squeezed baths requires
the consumption of external work.

Next, we use similar arguments to those in \citep{Wolgang2018passivityCI}
to show that squeezing can be fully captured in a slightly modified
version of the CI. For this, we need to introduce the notion of passive
states, ergotropy, and passive energy \citep{Wolgang2018passivityCI}.
A passive state (passive density matrix) with respect to a Hamiltonian
is one in which lower energy states are more populated than higher
energy states. In addition, the passive density matrix has no coherences
in the energy basis of the Hamiltonian. These states are called passive
states since no transient local unitary (acting on the system alone)
can reduce the average energy of a system prepared in this state.
That is, no work can be extracted from such a state if at the end
of the process the system Hamiltonian returns to its initial value
(transient unitary). Passive states are discussed in detail in Sec.
\ref{subsec: Traditional-passivity} and \ref{subsec: Global-passivity}. 

The maximal amount of work that can be extracted using local transient
unitaries is called ergotropy \citep{AllahverdyanErgotropy}. It is
obtained by bringing a non-passive distribution into its passive form
with respect to the Hamiltonian. Let $\mathcal{E}$ denote the 'passive
energy' \citep{Wolgang2018passivityCI} that is defined as the energy
that remains after extracting all the ergotropy. It is the energy
of the passive state associated with some initial non-passive state.
Since unitaries do not change the eigenvalues of the density matrix,
the initial non-passive state dictates the eigenvalues of the passive
state. Therefore, the initial state uniquely determines the passive
state up to degeneracies in the Hamiltonian. These possible degeneracies
are not important in the present chapter.

The entropic form of the CI (\ref{eq: DS^j}) holds regardless of
squeezing, since squeezing is a local operation that does not change
the local entropies. Applying unitary invariance of the VN entropy
to squeezed microbaths we can write 
\begin{equation}
\Delta S^{\mu b}=S(\rho_{f}^{\mu b})-S(\rho_{sq}^{\mu b}),
\end{equation}
as 
\begin{align}
\Delta S^{\mu b} & =S(U_{f}\rho_{fin}^{\mu b}U_{f}^{\dagger})-S(U_{i}\rho_{sq,\beta}^{\mu b}U_{i}^{\dagger})\nonumber \\
= & S(\rho_{f,pass}^{\mu b})-S(\rho_{\beta}^{\mu b}),
\end{align}
where $U_{f}$ is the unitary that takes the final state of the microbath
into a passive state, and $U_{i}$ is the unitary that takes the initial
state of the microbath to a thermal passive state. Using (\ref{eq: DS^j})
and (\ref{eq: basic DS identity}) for $\rho_{f,pass}^{\mu b}$ and
$\rho_{\beta}^{\mu b}$ we obtain the passive form of the CI \citep{Wolgang2018passivityCI}

\begin{equation}
\Delta S^{sys}+\sum\beta_{k}\Delta\mathcal{E}_{k}\ge0,\label{eq: passive CI}
\end{equation}
where $\beta_{k}$ is the initial temperature of the unsqueezed $k$-th
microbath, and $\mathcal{E}_{k}$ is the passive energy of the $k$-th
microbath. One can argue that $\Delta\mathcal{E}_{k}$ is a more accurate
definition of heat compared to $\Delta\left\langle H_{k}\right\rangle $
as it excludes ergotropy. However, another point of view is that heat
is related to energy that cannot be operationally extracted \textit{in
practice}. If retrieving the ergotropy from the bath is too complicated
to implement then for all practical purposes all energy dumped into
the microbath can be considered as heat. For macroscopic baths, direct
work on the bath is usually not considered as it may often involve
keeping track of phases of many interacting degrees of freedom. However,
in the microscopic world where the size of the microbath may be comparable
to that of the system, work extraction from a microbath is not an
unreasonable operation. In conclusion, the decision whether to use
(\ref{eq: basic CI}) or (\ref{eq: passive CI}) may strongly depend
on the experimental capabilities in a specific setup. 

That said, one of the nice features of the CI (\ref{eq: basic CI})
is that it does not depend on heat and work separation or interpretation.
It simply puts a limit on the changes in the average energy of the
microbaths. Work and heat separation and the very definition of work
becomes very obscure, and to some extent immaterial, when trying to
define higher moments of work in the presence of initial coherences
in the energy basis (see, for example, Ref. \citep{Marti2017NoGoQWorkFluch}).
On the other hand, changes in higher moments of the energy of the
system are well defined even in the presence of initial coherences
(see Sec. \ref{subsec: FT vs CI}). This is exploited in \citep{RU2017genCI}.
See \citep{Wolgang2018passivityCI,BeraWinters2ndLawCat} for schemes
that suggest some alternative separations of heat and work.

\subsection[Quantum coherence and the CI]{Quantum coherence and the Clausius inequality\label{subsec: Quantum-coherence}}

The microscopic form of the CI (\ref{eq: basic CI}) is so similar
to the historical macroscopic form put together by Clausius (\ref{eq: historic CI}),
that one may wonder if (\ref{eq: basic CI}) contains any quantum
features at all. The difference between stochastic dynamics of energy
population and quantum dynamics manifests in 'coherences': the off-diagonal
elements of the density matrix that quantify quantum superposition
of energy states. Coherence and its quantification is actively studied
in recent years\citep{PlenioAdesso2017coherenceRev}.

Removal of the system coherence by some process increases the entropy
of the system. According to the CI the growth of entropy sets a bound
on how much heat can be exchanged in such a decoherence process that
affects only the off diagonal elements of the density matrix. Indeed,
as discussed in Sec. \ref{subsec: Two-deficiencies} by using a specific
type of system-environment dephasing interactions it is possible to
decohere the system without changing the energy of the system or the
microbath (no heat flow). This thermodynamically irreversible process
involves significant correlation buildup between the system and the
microbath \citep{GlobalPassivity}. However, in Ref. \citep{Anders2015MeasurementWork}
a protocol was suggested to remove the coherence in a reversible process.

The Kammerlander-Anders protocol \citep{Anders2015MeasurementWork}
reversibly implements $\rho_{i}\to\text{diag}(\rho_{i})$ in two stages.
In stage $I$, a unitary transformation is applied to the system,
and brings it into an energy diagonal state. This step involves work
without any entropy changes $S_{I}'=S_{I}=S(\rho_{i})$. In stage
$II$ a standard reversible diagonal/stochastic state preparation
protocol is employed to change $\text{diag}(\rho')$ to $\text{diag}(\rho_{i})$.
The entropy change in this stage is
\begin{equation}
\Delta S_{II}=S[\text{diag}(\rho_{i})]-S_{I}'=S[\text{diag}(\rho_{i})]-S(\rho_{i}).
\end{equation}
Note that this expression is always positive as it can be written
in terms of quantum relative entropy

\begin{equation}
\Delta S_{II}=S[\text{diag}(\rho_{i})]-S(\rho_{i})=D[\rho_{i}|\text{diag}(\rho_{i})]\ge0.
\end{equation}
The heat required to reversibly generate this entropy difference using
a single-bath is given by the CI
\begin{equation}
q^{rev}=-T\Delta S_{II}.
\end{equation}
Since the initial and final average energy are the same (same values
of the diagonal elements, and the same Hamiltonian) it follows that
the work in this reversible process is
\begin{equation}
W^{rev}=q^{rev}=TD[\rho_{i}|\text{diag}(\rho_{i})]\ge0.
\end{equation}
Since the CI dictates $q\ge q^{rev}$ it follows that $W^{rev}\ge W$.
This result means that work can always be extracted by reversibly
removing the coherence in a reversible way. Moreover, reversible protocols
extract the maximal amount of work from coherences. Using the CI it
was not necessary to describe the protocol in full detail (see \citep{Anders2015MeasurementWork,RU2017genCI})
to obtain the optimal work extraction. 

Another way to think of the thermodynamic role of coherence in the
energy basis is to define the non-equilibrium free energy \citep{CrooksThemoPredNonEf,Esposito2011EPL2Law}
$\mc F$ through the Clausius inequality (\ref{eq: basic CI}). Starting
with the CI

\begin{equation}
\Delta(S+\beta\Delta\left\langle H^{\mu b}\right\rangle )\ge0,
\end{equation}
and using energy conservation $W+\Delta\left\langle H^{\mu b}\right\rangle +\Delta\left\langle H^{sys}\right\rangle =0$
(no initial or final interaction terms), we get the non-equilibrium
free energy form of the CI
\begin{align}
W & \le-\Delta\mc F,\label{eq: W DF CI}
\end{align}
where $\mc F=U-TS$ is the non-equilibrium free energy. As in the
CI, (\ref{eq: W DF CI}) becomes equality for reversible processes.
Since $S[diag(\rho^{sys})]\ge S(\rho^{sys})$, (\ref{eq: W DF CI})
predicts that in a reversible coherence removal process the extracted
work is $W=-T\{S(\rho^{sys})-S[diag(\rho^{sys})]\}>0$ as obtained
by the KA protocol described above.

Based on the derivations in the last two sections, we can add two
more items to CI properties listed in Sec. \ref{sec: CI in micro setups}
\begin{itemize}
\item The passive CI can handle squeezed thermal baths.
\item System coherences are taken into account, and play an important role
in the microscopic CI. 
\end{itemize}

\section{Global passivity approach\label{sec: Global-passivity-approach}}

\subsection{Traditional passivity and work extraction\label{subsec: Traditional-passivity}}

Passivity was broadly used in thermodynamics in the context of work
extraction \citep{pusz78,lenard1978Gibbs,AllahverdyanErgotropy,Wolgang2018passivityCI}.
We start with the definition of passivity, and then exploit it in
new ways \citep{GlobalPassivity}. Consider a system subjected to
a transient pulse
\[
H(t)=H_{0}+H_{pulse}(t).
\]
The pulse satisfies $H_{pulse}(t\le0;t\ge\tau_{pulse})=0$ so after
time $\tau_{pulse}$, the Hamiltonian returns to its initial form.
However the final density matrix is modified $\rho_{f}\neq\rho_{0}$
due to non-adiabatic coupling the pulse induced. Since this Hamiltonian
generates a unitary transformation, all eigenvalues of the initial
density matrix $\lambda(\rho_{0})$ must be conserved (in particular,
the entropy is conserved). To maximize the amount of average energy
the pulse is extracting from the system (work) we want to minimize
the quantity 
\begin{equation}
\Delta\text{\ensuremath{\left\langle H_{0}\right\rangle }=}\text{tr}[\rho_{f}H_{0}]-\text{tr}[\rho_{0}H_{0}].
\end{equation}
The second term is fixed by the initial condition, so the first term
$\text{tr}[\rho_{f}H_{0}]$ has to be minimized. Since the eigenvalues
are conserved $\lambda(\rho_{f})=\lambda(\rho_{0})$, the minimal
value of $\text{tr}[\rho_{f}H_{0}]$ is obtained by making $\rho_{f}$
diagonal in the energy basis and assigning lower energies with higher
eigenvalues of the density matrix (probabilities). This is called
the passive distribution. Writing $\lambda_{1}^{\downarrow}\ge\lambda_{2}^{\downarrow}\ge\lambda_{3}^{\downarrow}...$
the passive density matrix with respect to the Hamiltonian is
\[
\rho_{pass}=\sum_{k}\lambda_{k}^{\downarrow}\ketbra{k^{\uparrow}}{k^{\uparrow}}
\]
where $\ket{k^{\uparrow}}$ are the eigenstates of $H_{0}$ sorted
in energy-increasing order $E_{k}\le E_{k+1}\le E_{k+2}...$ Alternatively
stated, for a system starting in a passive state $\rho_{pass}$ with
respect to $H_{0}$ it holds that
\begin{equation}
\Delta\left\langle H_{0}\right\rangle _{pass\to fin}=\text{tr}[\rho_{f}H_{0}]-\text{tr}[\rho_{pass}H_{0}]\ge0,\label{eq: D H pass}
\end{equation}
for any transient unitary (pulse-like operation). Moreover, from linearity,
(\ref{eq: D H pass}) also holds for any mixture of unitaries (\ref{eq: mix of uni}).
As mentioned earlier, the maximal amount of work that can be extracted
from an initial state using a transient unitary process is called
ergotropy \citep{AllahverdyanErgotropy} and is equal to
\[
erg=-(\text{tr}[\rho_{pass}H_{0}]-\text{tr}[\rho_{0}H_{0}])\ge0.
\]

Thermal states are passive with respect to the Hamiltonian, but not
all passive states are thermal. It turns out that the relationship
between thermal states and passive states has a more profound aspect.
For a given Hamiltonian $H_{0}$ there are many possible passive states
(if $\rho_{0}$ is not specified as it was above). However, thermal
states are the only \textit{completely passive} states. That is, when
taking any number $n$ of uncorrelated copies in a thermal state $\rho_{tot}=\rho_{\beta}^{\otimes n}$
then $\rho_{tot}$ is always passive with respect to the total Hamiltonian
$H_{0}\otimes I\otimes I..+I\otimes H_{0}\otimes I...+...$ ($I$
is the identity operator). In particular, the thermal state is the
only state that remains passive in the limit $n\to\infty$. This is
a very important property from the point of view of the second law.
If the thermal state was not completely passive then by joining many
identical thermal microbaths we could have obtained a non-passive
state, and extract work from it, which is in contradiction to the
second law.

\subsection{Global passivity\label{subsec: Global-passivity}}

It is important to understand that passivity is not a property of
a state $\rho$ or of an operator $\mathcal{A}$, it is a joint property
of the pair $\{\rho,\mathcal{A}\}.$ While passivity was traditionally
used to describe work extraction of from a system, in \citep{GlobalPassivity}
it was suggested that passivity is a much more general concept that
can be used for other operators (not just Hamiltonians) and to other
objects (not just the system). Based on these observation it was shown
in \citep{GlobalPassivity} that passivity plays a much more significant
role in thermodynamics compared to the way it was used thus far. 

First, we point out that there is no reason to limit passivity to
be with respect to the Hamiltonian of the system or the bath. The
definition above is applicable to any observable described by an Hermitian
operator. For example given a state $\rho_{0}$ in some basis one
can ask what is the passive state with respect to the angular momentum
operator. In what follows, not only that we will not use passivity
with respect to the Hamiltonian, we will also use observables (operators)
that involve the \textit{whole} setup.

Second, usually in the context of passivity, $\mathcal{A}$ is given
(the Hamiltonian of the system or of a microbath), and the focus is
on the passive state associated with it \citep{pusz78,lenard1978Gibbs,AllahverdyanErgotropy,Marti2015EnergeticPassive,MartiWorkCorr,Wolgang2018passivityCI}.
In \citep{GlobalPassivity} it was suggested to look on the opposite
problem. Given an initial state (which may not be passive with respect
to the Hamiltonian), what are the passive operators with respect to
this state? That is, \textit{what are the $\mathcal{A}$'s such that
$\Delta\left\langle \mathcal{A}\right\rangle \ge0$ for any (transient)
unitary transformation }(\ref{eq: uni evo})\textit{ or a mixture
of unitaries} (\ref{eq: mix of uni})? 

Let $\rho_{0}^{tot}$ be the density matrix that describes the total
setup (system plus microbaths). A thermodynamic protocol is a sequence
of unitary operations that describe system-environment interaction
and interaction with external fields (e.g. a laser field) that act
as work repositories. The accumulated effect of this protocol is given
by a global unitary $U$ or by a mixture of unitaries (\ref{eq: mix of uni}). 

Global passivity \citep{GlobalPassivity} is defined as follows: An
operator $\mc B$ is globally passive (with respect to $\rho_{0}^{tot}$),
if it satisfies
\begin{equation}
\Delta\left\langle \mc B\right\rangle \ge0,
\end{equation}
for any thermodynamic protocol (for any $p_{k}$, and $U_{k}$ in
(\ref{eq: mix of uni})). 

As we shall see for any initial preparation $\rho_{0}^{tot}$ there
are many families of globally passive operators. To systematically
find the passive operators associated with $\rho_{0}^{tot}$ we can
use the fact that in thermodynamic setup $\rho_{0}^{tot}$ is explicitly
known (in contrast to $\rho_{f}^{tot}$) and construct operators from
it. The simplest choice with strongest kinship to the CI is\textbf{
\begin{equation}
\mc B^{tot}=-\ln\rho_{0}^{tot}.\label{eq: Btot def}
\end{equation}
}We emphasize that this is a \textit{time-independent operator} and
it is linear in the instantaneous density matrix $\rho_{t}^{tot}$.
That is, the expectation value at time $t$ is $\left\langle \mc B^{tot}\right\rangle _{t}=\text{tr}[\rho_{t}^{tot}(-\ln\rho_{0}^{tot})]$.
It is easy to verify that this operator is globally passive. According
to (\ref{eq: Btot def}) the probability of observing an eigenvalue
$\lambda_{\mc B}$ is $p_{\lambda_{\mc B}}=e^{-\lambda_{\mc B}}$.
Hence, larger eigenvalues are associated with lower probabilities
and we can conclude that $\rho_{0}^{tot}$ and $\mc B^{tot}$ form
a passive pair, and therefore
\begin{equation}
\Delta\left\langle \mc B^{tot}\right\rangle \ge0,\label{eq: Btot ineq}
\end{equation}
for any mixture of unitaries (\ref{eq: mix of uni}). The connection
of the global passivity inequality (\ref{eq: Btot ineq}) to the standard
CI will be explored in the next section. However, a major difference
already stands out: in contrast to the CI (\ref{eq: Btot ineq}) holds
for any initial $\rho_{0}^{tot}$ even if the system and microbaths
are all initially strongly correlated. Thus, (\ref{eq: Btot ineq})
has the potential to go beyond a mere re-derivation of the CI.

\subsection{The observable-only analog of the CI}

To see the first connection between (\ref{eq: Btot ineq}) and the
CI we assume the standard thermodynamics assumption on the initial
preparation (\ref{eq: basic r0}) and get that relation (\ref{eq: Btot ineq})
now reads

\begin{align}
\Delta\left\langle \mc B^{sys}\right\rangle +\sum_{k}\beta_{k}q_{k} & \ge0,\label{eq: Btot inequ explicit}
\end{align}
where $\mc B^{sys}$ is a time-independent operator 
\begin{equation}
\mc B^{sys}\doteq-\ln\rho_{0}^{sys},\label{eq: Bsys def}
\end{equation}
and as in Sec. \ref{subsec:Energy-information-form} $q_{k}=\Delta\left\langle H_{k}\right\rangle $
is the change in the average energy of the $'k'$-th bath. Form (\ref{eq: Btot inequ explicit})
is linear in the final density matrix and involves only expectations
values. Equation (\ref{eq: Btot inequ explicit}) looks similar to
the CI (\ref{eq: basic CI}), but instead of the change in the entropy
$\Delta S^{sys}$, there is a change in the expectation value of the
operator $\mc B^{sys}.$ Before doing the comparison it is important
to point out that in cases we have only microbaths and no system (e.g.
in absorption refrigerator tricycles \citep{k272}, or in a simple
bath-to-bath heat flow) then both (\ref{eq: Btot inequ explicit})
and the CI reduce to $\sum_{i}\beta_{i}q_{i}\ge0$. To quantitatively
compare the standard CI and (\ref{eq: Btot inequ explicit}) in the
general case, we use (\ref{eq: basic DS identity}) to rewrite (\ref{eq: Btot inequ explicit})
as
\begin{equation}
\Delta S^{sys}+\sum_{k}\beta_{k}q_{k}\ge-D(\rho_{f}^{sys}|\rho_{0}^{sys}).\label{eq: Btot with DS}
\end{equation}
The term on the right-hand side is negative, which means that (\ref{eq: Btot inequ explicit})
is a weaker inequality compared to the CI. Nonetheless, (\ref{eq: Btot inequ explicit})
has an important merit. To experimentally evaluate $\Delta S^{sys}$
a full system tomography is needed ( $S_{f}^{sys}$ is calculated
from $\rho_{f}^{sys}$). In contrast, for $\Delta\left\langle \mc B^{sys}\right\rangle $
we need to measure only the expectation value of $\mc B^{sys}$. Not
only that this involves only $N$ elements out of the full $N\times N$
density matrix of the system, the elements (probability in the basis
of $\mc B^{sys}$) need not be known explicitly. The average $\left\langle \mc B^{sys}\right\rangle $
converges much faster compared to evaluation/estimation of the individual
probabilities via tomography.

\subsection[CCI - Correlation compatible CI]{CCI - Correlation compatible Clausius inequality\label{subsec: CCI}}

To get the full energy-information form of the CI (\ref{eq: basic CI}),
and to \emph{go beyond} the standard validity regime (\ref{eq: basic r0}),
we introduce the notion of strong passivity-divergence relation \citep{GlobalPassivity}.
We start by writing the identity (\ref{eq: basic DS identity}) for
the whole setup
\begin{equation}
\Delta\left\langle \mc B^{tot}\right\rangle \equiv\Delta S^{tot}+D(\rho_{f}^{tot}|\rho_{0}^{tot})\label{eq: setup B ident}
\end{equation}
where $\mc B^{tot}$ is defined in (\ref{eq: Btot def}). When a specific
protocol described by a global unitary $U$ is applied to the setup,
then $\Delta S^{tot}=0$ since the eigenvalues of the total density
matrix do not change in a unitary evolution. What if we have some
noise in our controls that implements $U$ and the evolution is described
by a mixture of unitaries? In this case it holds that 
\begin{equation}
\Delta S^{tot}\ge0.\label{eq: DStot}
\end{equation}
This result can be obtained from the concavity of the von Neumann
entropy:
\begin{align}
S(\rho_{f}^{tot}) & =S(\sum p_{k}U_{k}\rho_{0}^{tot}U_{k}^{\dagger})\ge\nonumber \\
 & \sum_{k}p_{k}S(U_{k}\rho_{0}^{tot}U_{k}^{\dagger})=S(\rho_{0}^{tot}).
\end{align}
Property (\ref{eq: DStot}) is not unique to the von Neumann entropy,
it holds for any Schur concave function \citep{Marshall1979MajorizationBook}.
A density matrix created by a mixture of unitaries ($\rho_{f}^{tot}$)
is majorized by the initial density matrix $\rho_{f}^{tot}\prec\rho_{0}^{tot}$.
Therefore, for any Schur concave function $\mc S$, it holds that
$\mc S$($\rho_{f}^{tot})\ge\mc S(\rho_{0}^{tot})$ \citep{Marshall1979MajorizationBook}.
Equation (\ref{eq: DStot}) should not be confused with (\ref{eq: DS^j})
that describes the increase of the sum of reduced entropies when starting
from uncorrelated state. While (\ref{eq: DS^j}) contains the essence
of the standard CI, (\ref{eq: DStot}) describes another layer of
irreversibility created by the noise in the protocol. As we shall
see shortly the CI will be obtained from (\ref{eq: setup B ident})
even when there is no randomness in the protocol and $\Delta S^{tot}=0$.

Using (\ref{eq: DStot}) in (\ref{eq: setup B ident}), we obtain
the 'passivity-divergence relation'
\begin{equation}
\Delta\left\langle \mc B^{tot}\right\rangle \ge D(\rho_{f}^{tot}|\rho_{0}^{tot}).\label{eq: strong passivity}
\end{equation}
This inequality can be viewed as a stronger version of passivity for
the following reasons: First, global passivity (\ref{eq: Btot ineq})
immediately follows from (\ref{eq: strong passivity}) due to the
non-negativity of the quantum relative entropy. Second, (\ref{eq: strong passivity})
implies that the change in the expectation value is not only non-negative,
but also larger than $D(\rho_{f}^{tot}|\rho_{0}^{tot})\ge0$. Equations
(\ref{eq: setup B ident})-(\ref{eq: strong passivity}) constitute
an alternative way of proving (\ref{eq: Btot ineq}). 

The passivity-divergence relation (\ref{eq: strong passivity}) is
expressed in terms of the setup states without any explicit reference
to the system. To obtain an energy-information form (system's information)
we use the following property of relative entropy 
\begin{equation}
D(\rho_{f}^{tot}|\rho_{0}^{tot})\ge D(\rho_{f}^{sys}|\rho_{0}^{sys}).\label{eq: Dtot Dsys}
\end{equation}
This property follows from joint convexity of the quantum relative
entropy and it holds for any $\rho_{f}^{tot},\rho_{0}^{tot}$ even
in the presence of quantum or classical correlations. Mathematically,
the quantum relative entropy is a \textit{divergence}. Divergence
is a measure that quantifies how different two density matrices are.
In general, it is not a distance in the mathematical sense as it may
not satisfy the triangle inequality. Equation (\ref{eq: Dtot Dsys})
states that the disparity of the reduced states is smaller than the
disparity of the total states. This is plausible since some of the
differences are traced out. Nevertheless, there are divergences which
do not satisfy this property. Using (\ref{eq: strong passivity}-\ref{eq: Dtot Dsys})
we get 
\begin{equation}
\Delta\left\langle \mc B^{tot}\right\rangle \ge D(\rho_{f}^{sys}|\rho_{0}^{sys}).\label{eq: Btot Dsys}
\end{equation}
As we show next it is this very step that generates an extended version
of the CI. By applying (\ref{eq: basic DS identity}) to the relative
entropy of the system in (\ref{eq: Btot Dsys}) we get

\begin{equation}
\Delta\left\langle \mc B^{tot}\right\rangle \ge-\Delta S^{sys}+\Delta\left\langle \mc B^{sys}\right\rangle .
\end{equation}
where $\mc B^{sys}$ is defined by (\ref{eq: Bsys def}) even if the
system is initially correlated to the environment. Rearranging we
obtain the \textit{correlation compatible Clausius inequality} (CCI)
\citep{GlobalPassivity}

\begin{equation}
\Delta S^{sys}+\Delta\left\langle \mc B^{tot}\right\rangle -\Delta\left\langle \mc B^{sys}\right\rangle \ge0,\label{eq: CCI}
\end{equation}
For initially uncorrelated system and environment $\rho_{0}^{tot}=\rho_{0}^{sys}\otimes\rho_{0}^{env}$
it hold that $\mc B^{tot}=\mc B^{sys}\otimes I^{env}+I^{sys}\otimes\mc B^{env}$
and the CCI reduces to the CI (\ref{eq: CI Bbath})
\begin{align}
\Delta S^{sys}+\Delta\left\langle \mc B^{env}\right\rangle  & \ge0.
\end{align}
Therefore, we conclude that in the presence of initial correlations
the environment operator $\mc B^{env}$ must be replaced by $\mc B^{tot}-\mc B^{sys}$.
This is the content of the CCI. The local environment expectation
value is replaced by a global expectation value. If the environment
is composed of microbaths initially prepared in a thermal state we
get the standard CI (\ref{eq: basic CI}). Note that the CCI can also
be written as
\begin{equation}
\Delta S^{sys}+\Delta\left\langle \mc B^{env}\right\rangle +\Delta\left\langle \mc B^{corr}\right\rangle \ge0,
\end{equation}
where
\begin{equation}
\mc B^{corr}=\mc B^{tot}-\mc B^{sys}\otimes I^{env}-I^{sys}\otimes\mc B^{env},
\end{equation}
and $I^{sys},I^{env}$ are identity operators. The correlation operator
$\mc B^{corr}$ becomes identically zero where $\rho_{0}^{tot}=\rho_{0}^{sys}\otimes\rho_{0}^{env}$.
Note that $\left\langle \mc B^{corr}\right\rangle _{t_{0}}$ is a
measurable correlation quantifier of the initial state as it contains
only expectation values (in contrast to mutual information, for example). 

\subsubsection*{The CCI for a coupled system-environment initial thermal state}

Next, we consider an important case where initial correlations naturally
arise. In the setup in Fig \ref{fig: init_setups}b there is a cold
microbath and a hot microbath and the system is initially coupled
to the hot microbath. In the present scenario, we assume the system
was coupled to the hot microbath while the hot microbath was prepared
(e.g., by weak coupling to a macroscopic bath). As a result, the hot
microbath is not in a thermal state, rather, the microbath plus the
system are in a thermal state. Hence, the initial density matrix is
\begin{equation}
\rho_{0}^{tot}=\frac{1}{Z_{hs}Z_{c}}e^{-\beta_{h}(H_{h}+H_{s}+H_{int,0})}e^{-\beta_{c}H_{c}}.\label{eq: r0 coupled Gibbs}
\end{equation}
Using this in the CCI (\ref{eq: CCI}) we get:
\begin{equation}
\Delta S^{sys}+\beta_{c}q_{c}+\beta_{h}q_{h}+\beta_{h}\Delta\left\langle H_{int,0}\right\rangle +\beta_{h}(H_{s}-H_{s}^{eff})\ge0,\label{eq: CCI couple gibbs}
\end{equation}
where the effective Hamiltonian is defined via $\rho_{0}^{sys}=e^{-\beta H_{s}^{eff}}/Z$.
The $Z$ normalization factor yields an additive constant that can
omitted so 
\begin{equation}
H_{sys}^{eff}\doteq-\frac{1}{\beta_{h}}\ln\rho_{0}^{sys}=-\frac{1}{\beta_{h}}\ln tr_{h}e^{-\beta_{h}(H_{s}+H_{int,0}+H_{h})}.\label{eq: Heff sys}
\end{equation}
Note that all the Hamiltonians on the right hand side of (\ref{eq: Heff sys})
refer to their value at time zero, and not to their possibly different
instantaneous values. The term $H_{s}^{eff}-H_{s}$ is known as the
potential of mean force \citep{kirkwood1935PMF} or the solvation
Hamiltonian \citep{Jarzynski2017_PRX_strong_coupling}. The first
three terms in (\ref{eq: CCI couple gibbs}) are the standard ``bare''
Clausius terms. The fourth term represents changes in the interaction
energy and the last term is a system dressing effect. The last term
represents the fact that the reduced state of the system is not the
thermal state of the bare Hamiltonian of the system. See \citep{GlobalPassivity}
for an explicit calculation of $H_{sys}^{eff}$ for a dephasing interaction
and a swap interaction. 

The coupled thermal CCI (\ref{eq: CCI couple gibbs}) has some similarity
to a classical result \citep{Jarzynski2017_PRX_strong_coupling} in
a similar scenario. However there are two important differences. First,
our result is valid also in the presence of coherences (in the energy
basis) and quantum correlations that arise from the initial system-environment
coupling. Second, the result in \citep{Jarzynski2017_PRX_strong_coupling}
is obtained from fluctuation theorems \citep{SeifertPRL2016_strong_coupling,JanetPRE2017_strong_coupling}.
As such, it involves the \textit{equilibrium} entropy (leading to
the and \textit{equilibrium} free energy). Our result involves the
von Neumann entropy (leading to the and \textit{non-equilibrium} free
energy \citep{CrooksThemoPredNonEf,Esposito2011EPL2Law}). 

\subsection[Global passivity outlook]{An outlook for the global passivity approach\label{subsec: GP Outlook}}

So far, we have used global passivity to extend the validity regime
of the CI to the case of initial system-environment correlation. In
doing so, we maintained the CI structure described in Sec. \ref{subsec: CI-Structure}.
Remarkably, global passivity can generate additional thermodynamic
inequalities that involve different quantities. In \citep{GlobalPassivity}
it was pointed out that since $(\mc B^{tot})^{\alpha}$ and $\mc B^{tot}$
have the same eigenvectors and the same eigenvalue ordering for $\alpha>0$
($(\mc B^{tot})^{\alpha}$ is a stretched/squeezed version of $\mc B^{tot}$)
it follows that $(\mc B^{tot})^{\alpha}$ is also globally passive
(with respect to $\rho_{0}^{tot}$) and therefore
\begin{equation}
\Delta\left\langle (\mc B^{tot})^{\alpha}\right\rangle \ge0,
\end{equation}
for any $\alpha>0$ and for any thermodynamic protocol (\ref{eq: mix of uni}).
These inequalities involve higher moments of the energy. In \citep{GlobalPassivity}
it was exploited to detect ``lazy Maxwell's demons'' (subtle feedback
operations) and hidden heat leaks that the CI cannot detect. Hopefully,
such inequalities will significantly extend the scope of the microscopic
thermodynamic framework as discussed in Sec. \ref{sec: Outlook-and-challenges}.

\section[CPTP maps approach]{CPTP maps approach}

The previous derivation is based on global arguments that take into
account the whole setup including the microbaths. In this section,
we adopt a system point of view where the environment is represented
by its action on the system. To simplify the notations, we use here
a single bath.

Using the definition of the relative entropy and the von Neumann entropy
it is straightforward to verify that the following identity holds
for any $\rho_{1},\rho_{2},\rho_{ref}$

\begin{equation}
S(\rho_{2})-S(\rho_{1})+tr[(\rho_{2}-\rho_{1})(\ln\rho_{ref})]\equiv D(\rho_{1}|\rho_{ref})-D(\rho_{2}|\rho_{ref}).\label{eq: 3RelEnt}
\end{equation}
By taking $\rho_{ref}=\rho_{1}$ in (\ref{eq: 3RelEnt}), (\ref{eq: basic DS identity})
is obtained. Alternatively, by using (\ref{eq: basic DS identity})
once for $\rho_{2},\rho_{ref}$ and once for $\rho_{1},\rho_{ref}$
it is possible to get (\ref{eq: 3RelEnt}). Focusing on the left hand
side we see a familiar structure, entropy difference followed by a
change in the expectation value of the Hermitian operator $\ln\rho_{ref}$.
Next we choose $\rho_{2}=\rho_{t}^{s},\rho_{1}=\rho_{0}^{s},\rho_{ref}=\rho_{\beta}^{s}$
and get

\begin{equation}
\Delta S_{sys}-\beta\Delta\left\langle H_{s}\right\rangle \equiv D(\rho_{0}^{sys}|\rho_{\beta})-D(\rho_{f}^{sys}|\rho_{\beta}).\label{eq: CI ident}
\end{equation}
Despite the close similarity to the CI expression, (\ref{eq: CI ident})
is an identity void of any physical content. Our first step, then,
is to assign a physical scenario to the left hand side. In order to
identify the change $\beta\Delta\left\langle H_{s}\right\rangle $
of the energy of the system with heat as in the CI, we need a scenario
with zero work. Using the standard system-based definitions of heat
and work 
\begin{align}
W & =\int^{t}\text{tr}[\rho_{s}\frac{d}{dt}H_{s}],\\
Q & =\int^{t}\text{tr}[H_{s}\frac{d}{dt}\rho_{s}],
\end{align}
we see that to have zero work, the Hamiltonian of the system has to
be fixed in time. Such a process is called an isochore since it is
the analog of the fixed volume process in macroscopic thermodynamics.
Thus, for an isochore it holds that

\begin{equation}
\Delta S_{sys}^{ISC}-\beta Q^{ISC}=D(\rho_{0}^{sys}|\rho_{\beta})-D(\rho_{f}^{sys}|\rho_{\beta}).
\end{equation}
The identity sign was removed as this expression holds just for isochores.
Now we are in position to focus on the right hand side.

\subsection[CPTP maps with a thermal fixed point]{Completely positive maps with a thermal fixed point}

Completely positive trace preserving maps (CPTP) are very useful in
describing measurements, feedback, dephasing, and interaction with
thermal baths. One way of representing CPTP maps is by using Kraus
maps. This is a local approach where the environment is not explicitly
described. Another way to describe a CPTP map is by interacting with
an auxiliary system (environment). Any CPTP map $\rho_{f}^{s}=M(\rho_{0}^{s})$
can be written as

\begin{equation}
\rho_{f}^{s}=\text{tr}_{A}[U\rho_{0}^{env}\otimes\rho_{0}^{s}U^{\dagger}],\label{eq: CPTP global}
\end{equation}
where $\rho_{0}^{env}$ is the initial density matrix of some environment
and $U$ is a global unitary that describes the system-environment
interaction. CPTP maps have the following monotonicity property for
any two density matrices $\rho,\sigma$ \citep{nielsen2002QuantInfo}
\begin{equation}
D[\rho|\sigma]\ge D[M(\rho)|M(\sigma)].\label{eq: gen contractivity}
\end{equation}
That is, a CPTP map is ``contractive'' with respect to the quantum
relative entropy divergence. Roughly speaking, operating with $M$
on two distinct states make them more similar to each other. If $M$
has a fixed point $M(\rho_{FP})=\rho_{FP}$, (\ref{eq: gen contractivity})
reads

\begin{equation}
D[\rho|\rho_{FP}]\ge D[M(\rho)|\rho_{FP}],\label{eq: FP contract}
\end{equation}
which means that when $M$ is applied all $\rho$'s ``approach''
the fixed point of the map. Next, we make the physical choice that
in isochores where a system is coupled to a bath, the thermal state
$\rho_{s}=e^{-\beta(H_{s}-F_{s})}$ is the fixed point. If we connect
an already thermal system to a bath in the same temperature, nothing
will happen. This is the CPTP equivalent of the zeroth law. This non-trivial
feature has to be properly justified and in what follows we discuss
the emergence of thermal fixed points in CPTP maps.

\subsubsection{Fixed points of CPTP maps}

Let our environment be initialized at a Gibbs state $\rho_{\beta}^{b}=e^{-\beta(H_{b}-F_{b})}/Z$.
Next, we assume that the interaction Hamiltonian $H_{int}$ commutes
with the total system-bath bare Hamiltonians
\begin{equation}
[H_{int},H_{s}+H_{b}]=0.\label{eq: Hint commute}
\end{equation}
This guaranties that $\left\langle H_{s}\right\rangle +\left\langle H_{b}\right\rangle =const$
and that no energy is transferred to the interaction energy. This
condition also assures that no work has to be invested in coupling
the system to the bath. In thermodynamic resource theory this condition
is often written as $[U,H_{s}+H_{b}]=0$. Under this condition, it
follows that the thermal state of the system is a fixed point of the
CPTP map since
\begin{align}
M(\rho_{\beta}^{s}) & =\text{tr}_{b}[U\rho_{\beta}^{b}\otimes\rho_{\beta}^{s}U^{\dagger}]\nonumber \\
 & =\text{tr}_{b}[Ue^{-\beta(H_{s}+H_{b}-F_{s}-F_{b})}U^{\dagger}]\nonumber \\
 & =\text{tr}_{b}[e^{-\beta(H_{s}+H_{b}-F_{s}-F_{b})}UU^{\dagger}]\nonumber \\
 & =\rho_{\beta}^{s}.\label{eq: CPTP FP}
\end{align}
There are other scenarios where fixed points occur. For example, if
two baths are coupled simultaneously to the same levels of the system,
the fixed point will not be thermal, in general. We call it a 'leaky
fixed point' since in such cases even in steady state there is a constant
heat flow (heat leak) between the baths. We will return to this point
at the end of the section.

\subsection{From fixed points to the Clausius inequality\label{subsec: fixed-point to CI}}

Under condition (\ref{eq: Hint commute}), $\rho_{\beta}^{s}$ is
a fixed point of a CPTP map obtained by interacting with an initially
thermal environment. Using the contractivity (\ref{eq: FP contract})
with respect to the thermal fixed point we get that for isochores
(ISC)
\begin{equation}
\Delta S_{ISC}^{sys}-\beta Q_{ISC}\ge0.
\end{equation}
To extend this isochore result to general thermodynamic scenarios
we need to include the possibility of doing pure work on the system
without contact with the bath. For a unitary transformation on the
system (this may include either changing the energy levels in time,
or applying an external field to change the population and coherences),
it holds that the heat flow is zero
\begin{align}
\frac{dQ}{dt} & =\text{tr}[\frac{d\rho^{s}}{dt}H(t)]=\text{tr}\{-i[H(t),\rho]H(t)\}=0,
\end{align}
where we used the cyclic property of the trace in the last transition.
Furthermore, the system entropy does not change under local unitaries,
hence for unitary evolution (UNI) $\Delta S_{UNI}^{sys}=0,Q_{UNI}=0$.
The unitary evolution is the analog of the macroscopic thermodynamic
adiabat. Now, if we have a sequence of an isochore and a unitary,
the Clausius term is

\begin{align}
\Delta S-\beta Q & =(\Delta S_{ISC}+\Delta S_{UNI})+\beta(Q_{ISC}+Q_{UNI})\\
 & =\Delta S_{ISC}+\beta Q_{ISC}\ge0.
\end{align}
Hence,
\begin{equation}
\Delta S-\beta Q\ge0,\label{eq: CIQ}
\end{equation}
for any concatenation of isochores (thermal CPTP maps) and unitaries.
One can show that other processes such as isotherms can be constructed
from a concatenation of isochores and adiabats \citep{anders2013thermodynamics,RU2017genCI}.

For isotherm, $\Delta S=\beta Q$ and the reversible saturation of
the CI is obtained. This can also be shown by direct integration.
For isotherms, the state is always in a thermal state even when the
Hamiltonian (slowly) changes in time. Therefore $\rho_{s}=e^{-\beta(t)[H(t)-F(t)]}$
and $H(t)=F(t)+T(t)\ln\rho^{s}(t)$ ($F$ is a scalar matrix). Using
it in the entropy definition we get

\begin{equation}
\Delta S=\int^{t}\text{tr}[-(\frac{d}{dt}\rho_{s})\ln\rho_{s}]=\int^{t}\text{tr}\{\beta(t)[H(t)-F(t)]\frac{d}{dt}\rho_{s}\}=\int^{t}\text{\ensuremath{\beta}}\frac{dQ}{dt}.
\end{equation}
Note that $Q$ represents the change in the energy of the system due
to the interaction of the bath. However, since we assumed that condition
(\ref{eq: Hint commute}) holds, it follows that $-Q=q=\Delta\left\langle H_{b}\right\rangle $.
Thus, we have retrieved the reversible saturation in an isothermal
reversible process (for unitaries it is trivially satisfied).

This derivation has two main drawbacks. The first is that condition
(\ref{eq: Hint commute}) does not always hold. In particular, in
short interaction time, before the rotating wave approximation becomes
valid, there are counter-rotating terms that do not satisfy (\ref{eq: Hint commute}).
The second drawback concerns the ability of CPTP maps to describe
a general interaction with a thermal bath. As mentioned earlier, a
general protocol can be decomposed into adiabats and isochores. However,
the assumption that at each isochore the fixed point is the thermal
state requires that the bath be large and the coupling to it is weak.
Weak coupling is needed in order to ensure negligible system-bath
correlation at the beginning of each isochore (see (\ref{eq: CPTP global})).
This is necessary for using the CPTP contractivity property (\ref{eq: gen contractivity})
and (\ref{eq: FP contract}).

Interestingly, in the presence of heat leaks, the local fixed point
approach may provide different predictions from the standard global
approach to the CI. There is no contradiction between the local approach
described above (\ref{eq: CIQ}) and the CI, but they provide predictions
on different quantities. Consider the case where a two-level system
is connected simultaneously to two large thermal baths with temperature
$1/\beta_{1}$ and $1/\beta_{2}$. The fixed point of the two-level
system will be some diagonal state with some intermediate temperature
$1/\beta_{eff}$ that depends on the interaction strength (thermalization
rate) with each bath. In systems with more levels, the steady state
will not be thermal in general. According to the local approach (\ref{eq: CIQ})
the $\beta$ that should be used in the CI is $\beta_{eff}$ so that
$\Delta S-\beta_{eff}Q\ge0$. However, according to the global approach
(\ref{eq: CI Bbath}), we have $\Delta S+\beta_{1}q_{1}+\beta_{2}q_{2}\ge0$.
Both expressions are correct but they contain information on different
quantities. 

\section{Outlook and challenges\label{sec: Outlook-and-challenges}}

New approaches to the second law, as well as new mathematical tools
such as those discussed in the previous sections (e.g. Sec. \ref{sec: Global-passivity-approach}),
lead to additional thermodynamic constraints on operations that involve
thermal environments. In the context of the second law in microscopic
setups, there are several main challenges that deserve further research:
\begin{enumerate}
\item Strongly correlated thermal systems
\item Correlation dynamics and higher order energy moments
\item X heat machines
\item Heat leaks and feedback detection
\item Deviation from the standard energy-information paradigm
\item Fluctuation theorems and CI extensions
\end{enumerate}

\subsection{Strongly correlated thermal systems}

Consider a system composed of several dozen interacting particles,
e.g., interacting spins in a lattice. The particles are initially
in a thermal state. Next, a unitary operation (e.g., lattice shaking,
or interactions with external fields) is applied to the setup and
takes it out of equilibrium. When the particle number exceeds three
dozen or so, the dynamics cannot be carried out numerically with present
computational resources. Hence, thermodynamic predictions can be quite
useful. Applying the second law to the whole setup yields a very trivial
result. There is no change in entropy, and there is no heat exchange
with some external environment. The energy changes in the setup are
purely work related. 

To understand the \textit{internal} energy flows and \textit{local
entropy} changes, we need to choose some artificial partitioning according
to our zone of interest. The zone of interest constitutes the 'system',
and everything else is the 'environment' (even if it is very small
and comparable to the system's size). For example, the system can
be a collection of a few neighboring spins or even a single spin.
The system can even be disconnected, e.g., two non-adjacent spins
or even a sparse lattice that includes only a subset of the total
number of spins (e.g., every second spin in a chain configuration).
In all these examples, it is interesting how the \textit{reduced}
entropy of the zone of interest changes (the total entropy of the
setup is conserved) when energy flows in and out of it. Due to the
equilibrium interaction, the spins are initially correlated to each
other. Consequently, the partitions suggested above cannot be studied
with the CI. In contrast, the CCI is suitable for this scenario. It
is interesting to explore what insights the CCI, (and other yet undiscovered
thermodynamics constraints), can provide to this important scenario.

\subsection[Correlations \& higher energy moments]{Correlation dynamics and higher order energy moments\label{subsec: Correlation and higher moments}}

The changes in the first moment of the microbath energy (heat) are
constrained by the CI. What about the second or even higher moments
of the energy of the microbath or the system? The CI does not deal
with these quantities. Does this mean that any change is possible,
or perhaps there are analogs of the CI that limit higher order moments
of the energy?

In large baths that to a good approximation remain in a thermal state
it is enough to know the changes in the average energy to predict
their final temperature. However, the final state of microbaths is
typically a non-equilibrium state. Thus, information on changes in
higher order moments of the energy becomes important. In \citep{RU2017genCI}
it was shown that in some scenarios it is possible to write Clausius-like
inequalities for higher moments of the energy. Interestingly, it is
shown that changes in higher-moments of the energy are associated
with information measure that differs from the von Neumann entropy.
If the new Clausius-like inequalities are tight (become equalities)
for reversible processes (as in \citep{RU2017genCI}), it is expected
that changes in non-extensive higher moments of the energy, be associated
with non-extensive information measures. The work in \citep{RU2017genCI}
is based on the local approach presented in Sec. \ref{subsec: fixed-point to CI}. 

Higher order moments of the energy are also very important when the
correlation buildup between system and bath is studied. The CI is
based on the fact that elements that start uncorrelated become correlated.
However, the CI does not set a bound on how big or how small these
correlations should be in terms of observable quantities. Consider
for example the system-bath covariance $\text{cov}(H_{s},H_{b})\doteq\left\langle H_{s}H_{b}\right\rangle -\left\langle H_{s}\right\rangle \left\langle H_{b}\right\rangle $.
When it is zero the system and bath are uncorrelated and when it equal
to $\sqrt{\text{var}(H_{s})}\sqrt{\text{var}(H_{b})}$ the system
and bath are maximally correlated. Unlike mutual information, the
covariance is an observable measure of correlation. However, it is
quadratic in energy, and therefore not constrained by the second law.
In \citep{GlobalPassivity} it was shown that in some cases global
passivity framework imposes constraints on the covariance buildup.
Moreover, system-environment correlations can develop between other
observables. For example, in a spin decoherence setup \citep{GlobalPassivity},
correlation builds up between the bath Hamiltonian and the initial
polarization operator of the spin. It is interesting to find thermodynamic
bounds on this covariance. The bounds in \citep{GlobalPassivity}
have not been proven optimal or unique. Thus, this topic warrants
further study.

\subsection{X machines\label{subsec: X-machines}}

Similarly to standard heat machines such as engine or refrigerators,
$X$ machines \citep{ruXmachines} are machines that exploit thermal
resources (e.g. microbaths) to execute a task of interest. However,
the task of the $X$ machine is not cooling (reduction of the entropy
or the average energy) or work extraction. The task of an X machines,
especially in the microscopic realm, can be much more fine-tuned and
customized for a specific scenario. Machines for entanglement generation
have been suggested and studied in \citep{HuberEntangGenHeat,Alexia2018steady}.

\begin{figure}
\includegraphics[width=10cm]{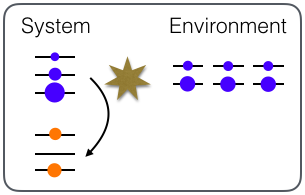}

\caption{\label{fig: Xmachine}An example of an 'X machine' setup. These machines
use thermal resources (a three-spin microbath in this case) to execute
a non-thermodynamic task such as the depletion of the second level
in the qutrit system. Since the goal of this type of machines is not
to manipulate average energy or entropy, the second law does not set
a performance bound for such machines.}
\end{figure}

As a concrete example of an X machine setup, consider a qutrit system
such as that shown in Fig. \ref{fig: Xmachine}. The system can interact
with a three-spin environment that is initially prepared in a thermal
state (microbath) with inverse temperature $\beta$. The goal of this
setup is to deplete as much as possible the second level of the qutrit.
This task is of importance when trying to inhibit an undesired interaction
or chemical reaction associated with the second level of the qutrit.
The task is carried out by some global unitary transformation on the
qutrit and spins setup. Crucially, to accomplish this goal, we allow
for both the entropy and the average energy of the qutrit to grow,
as long as the population of the second level reduces. Thus, this
task cannot be considered as cooling, heating, or work extraction.
More formally, the task of this heat machine is to minimize the expectation
value of a target operator $\mathcal{A}_{target}$ which in this case
equals to $\mathcal{A}_{target}=\ketbra{2_{sys}}{2_{sys}}$. 

On top of their potential practical value, X machines present us with
fantastic and exciting thermodynamic challenges. The minimization
of $\left\langle \mathcal{A}_{target}\right\rangle $ may be completely
unrelated to changes in the average energy or in the entropy. Thus,
even in cases where the CI holds for X machines, it does not provide
a performance bounds. For example, how well can these machines perform
as a function of the initial temperature of the environment and its
size? To understand this, it is vital to find additional CI-like inequalities
that will relate changes in $\left\langle \mathcal{A}_{target}\right\rangle $
(output), to changes in the microbath (resources). In analogy to reversible
processes in conventional machines, it will be very appealing to find
bounds that can be saturated by known protocols.  

\subsection[Heat leaks \& lazy demons detection]{Detecting heat leaks and lazy Maxwell's demons}

The CI has a clear regime of validity. If, for some reason, we find
that in our setup the CI does not hold, we can conclude that the process
in this setup is outside the regime of validity of the CI. The information
on being outside the regime of validity can be used to deduce some
conclusions on the cause of the anomaly. For example, in a Maxwell
demon setup, if we see heat flowing from the cold bath to the hot
bath (CI violation) without applying work, we can deduce the existence
of a Maxwell demon even if we do not observe the demon directly, only
the outcome of its operation. Similarly, in thermodynamic setups in
superconducting circuits, a violation of the CI may indicate that
the intrinsic thermalization to the background temperature cannot
be ignored. Thus, CI violation can provide information on the setup
and on the processes that take place.

Next, we ask how far it is possible to push this notion of detection
using the violation of thermodynamic inequalities. Perhaps the simplest
scenario to consider is the ``lazy Maxwell demon'' \citep{GlobalPassivity}.
The setup is initialized with cold molecules in one chamber and hot
molecules in another. A Maxwell demon that controls a trap door between
the chambers, measures the speed of the incoming molecules on both
sides, and according to the results, it decides whether to open or
close the door. If the demon performs properly, it can make the cold
bath colder and the hot bath hotter. This violation of the CI occurs
since feedback operations (the demon's action) are outside the regime
of validity of the CI (unless the feedback mechanism is included in
the setup \citep{leff2014MaxwellDemonBook}).

Consider the case where the demon is lazy, and it often dozes off
while the trap door is open. In these cases, the average energy (heat)
flows naturally from the hot bath to the cold bath. If the fraction
of the time when the demon is awake is too small, the cold bath will
get warmer, and the hot bath will get colder. Since in this case the
CI is not violated, the demon cannot be detected using the CI. We
ask if there are other thermodynamic inequalities that can detect
the presence of feedback even when the first moment of the energies
change consistently with the CI. In \citep{GlobalPassivity} it was
shown for a specific example that global passivity inequalities \citep{GlobalPassivity}
can detect lazy demons that the CI cannot detect. If, however, the
feedback is too weak (a very lazy demon), even the global passivity
inequalities in \citep{GlobalPassivity} may not be able to detect
it.

Although the findings in \citep{GlobalPassivity} show that this kind
of improved thermodynamic demon detection can exist, a big question
still remains: can any feedback operation, even a very subtle one,
be detected by some thermodynamic constraints on observables such
as the CI? Most likely, the resolution of this question will lead
to a more comprehensive thermodynamic framework.

\subsection[Deviation from the energy-info. relation]{Deviation from the standard energy-information paradigm}

In this chapter, an emphasis has been put on the energy-information
structure of the CI. However, in \citep{Goold2015nonEq_Landuer} a
fluctuation theorem has been used to derive an elegant relation between
average heat and a new type of measure $\mathcal{B}_{Q}$ that replaces
the entropy change of the system. The $B_{Q}$ measure is defined
as
\begin{equation}
B_{Q}=-\text{ln}tr(M\rho_{0}^{s}),
\end{equation}
where
\begin{equation}
M=tr_{E}[U(\rho^{env}\otimes I^{sys})U^{\dagger}],
\end{equation}
and $U$ stands for the global evolution operator. $M$ describes
how a system identity state ($I^{\ensuremath{sys}}/tr[I^{\ensuremath{sys}}]$,
super hot state), evolves under the operator $U$. The relation to
heat found in \citep{Goold2015nonEq_Landuer} is
\begin{equation}
\beta q\ge B_{Q}.
\end{equation}
 It is both interesting and important to understand the advantages
and disadvantages of $B_{Q}$ (and similar measures) with respect
to the CI. 

As mentioned in Sec. \ref{subsec: Two-deficiencies} one of the deficiencies
of the CI is its trivial prediction for extremely cold microbaths.
It simply predicts that energy in this super cold bath will increase.
This problem has been elegantly addressed for an harmonic oscillator
system using a phase space approach \citep{Paternostro2017EntProdWigner}.
For an initial Gaussian state, the authors find a CI-like relation
between the purity and the heat. In the CI-like expression in \citep{Paternostro2017EntProdWigner}
the diverging $\beta$ factor is replaced by an expression that does
not diverge as $\beta\to\infty$.

Another interesting deviation from the standard energy-information
employs catalysts and many copies of the same setup \citep{BeraWinters2ndLawCat}.
Using the theorem described in \citep{BeraWinters2ndLawCat} (see
also \citep{Carlo2017passive2thermal}), it is possible to employ
a global unitary and an external catalyst to change a given initial
density matrix $\rho_{0}$ to any isentropic state $\rho_{f}$ such
that $S(\rho_{f})=S(\rho_{0})$. Thus, it is possible to transform
the initial state to a thermal state $\rho_{\beta_{eff}}$, whose
temperature $1/\beta_{eff}$ is chosen to satisfy $S(\rho_{0})=S(\rho_{\beta_{eff}})$.
Now, all initial non-equilibrium scenarios can be treated as if they
are initially thermal with some effective temperature. For example,
if there is a non-thermal reservoir one can write a Clausius inequality
for it with $1/\beta_{eff}$ as its initial temperature. 

It is interesting to pursue these approaches and to find new approaches
that can overcome the deficiencies of the CI.

\subsection{Fluctuation theorems and CI extensions}

As mentioned in Sec. \ref{subsec: FT vs CI}, in their regime of validity,
fluctuation theorems (FT) yield stronger statements than the CI, and
they can reduce to the CI by using some mathematical inequalities
such as the Jensen inequality. As described in Sec. \ref{subsec: GP Outlook}
(Ref. \citep{GlobalPassivity}) and in Sec. \ref{subsec: Correlation and higher moments}
(Ref. \citep{RU2017genCI}) non-trivial extension of the CI can be
derived. These extensions provide inequalities on new observables
(e.g., higher order energy moments), and can extend the regime of
validity of the CI (e.g., to initial system-environment correlation). 

It is fascinating to investigate if some of these new CI extensions
can be obtained from fluctuation theorems.

\bibliographystyle{apsrev4-1}
\bibliography{/Users/raam_uzdin/Dropbox/RaamCite}

\begin{thebibliography}{53}%
\makeatletter
\providecommand \@ifxundefined [1]{%
 \@ifx{#1\undefined}
}%
\providecommand \@ifnum [1]{%
 \ifnum #1\expandafter \@firstoftwo
 \else \expandafter \@secondoftwo
 \fi
}%
\providecommand \@ifx [1]{%
 \ifx #1\expandafter \@firstoftwo
 \else \expandafter \@secondoftwo
 \fi
}%
\providecommand \natexlab [1]{#1}%
\providecommand \enquote  [1]{``#1''}%
\providecommand \bibnamefont  [1]{#1}%
\providecommand \bibfnamefont [1]{#1}%
\providecommand \citenamefont [1]{#1}%
\providecommand \href@noop [0]{\@secondoftwo}%
\providecommand \href [0]{\begingroup \@sanitize@url \@href}%
\providecommand \@href[1]{\@@startlink{#1}\@@href}%
\providecommand \@@href[1]{\endgroup#1\@@endlink}%
\providecommand \@sanitize@url [0]{\catcode `\\12\catcode `\$12\catcode
  `\&12\catcode `\#12\catcode `\^12\catcode `\_12\catcode `\%12\relax}%
\providecommand \@@startlink[1]{}%
\providecommand \@@endlink[0]{}%
\providecommand \url  [0]{\begingroup\@sanitize@url \@url }%
\providecommand \@url [1]{\endgroup\@href {#1}{\urlprefix }}%
\providecommand \urlprefix  [0]{URL }%
\providecommand \Eprint [0]{\href }%
\providecommand \doibase [0]{http://dx.doi.org/}%
\providecommand \selectlanguage [0]{\@gobble}%
\providecommand \bibinfo  [0]{\@secondoftwo}%
\providecommand \bibfield  [0]{\@secondoftwo}%
\providecommand \translation [1]{[#1]}%
\providecommand \BibitemOpen [0]{}%
\providecommand \bibitemStop [0]{}%
\providecommand \bibitemNoStop [0]{.\EOS\space}%
\providecommand \EOS [0]{\spacefactor3000\relax}%
\providecommand \BibitemShut  [1]{\csname bibitem#1\endcsname}%
\let\auto@bib@innerbib\@empty
\bibitem [{\citenamefont {Landauer}(1961)}]{Landauer1961InfoOrig}%
  \BibitemOpen
  \bibfield  {author} {\bibinfo {author} {\bibfnamefont {R.}~\bibnamefont
  {Landauer}},\ }\href@noop {} {\bibfield  {journal} {\bibinfo  {journal} {IBM
  journal of research and development}\ }\textbf {\bibinfo {volume} {5}},\
  \bibinfo {pages} {183} (\bibinfo {year} {1961})}\BibitemShut {NoStop}%
\bibitem [{\citenamefont {Reeb}\ and\ \citenamefont
  {Wolf}(2014)}]{reeb2014improved}%
  \BibitemOpen
  \bibfield  {author} {\bibinfo {author} {\bibfnamefont {D.}~\bibnamefont
  {Reeb}}\ and\ \bibinfo {author} {\bibfnamefont {M.~M.}\ \bibnamefont
  {Wolf}},\ }\href@noop {} {\bibfield  {journal} {\bibinfo  {journal} {New
  Journal of Physics}\ }\textbf {\bibinfo {volume} {16}},\ \bibinfo {pages}
  {103011} (\bibinfo {year} {2014})}\BibitemShut {NoStop}%
\bibitem [{\citenamefont {Brand{\~a}o}\ \emph {et~al.}(2015)\citenamefont
  {Brand{\~a}o}, \citenamefont {Horodecki}, \citenamefont {Ng}, \citenamefont
  {Oppenheim},\ and\ \citenamefont {Wehner}}]{BrandaoPnasRT2ndLaw}%
  \BibitemOpen
  \bibfield  {author} {\bibinfo {author} {\bibfnamefont {F.}~\bibnamefont
  {Brand{\~a}o}}, \bibinfo {author} {\bibfnamefont {M.}~\bibnamefont
  {Horodecki}}, \bibinfo {author} {\bibfnamefont {N.}~\bibnamefont {Ng}},
  \bibinfo {author} {\bibfnamefont {J.}~\bibnamefont {Oppenheim}}, \ and\
  \bibinfo {author} {\bibfnamefont {S.}~\bibnamefont {Wehner}},\ }\href@noop {}
  {\bibfield  {journal} {\bibinfo  {journal} {Proceedings of the National
  Academy of Sciences}\ }\textbf {\bibinfo {volume} {112}},\ \bibinfo {pages}
  {3275} (\bibinfo {year} {2015})}\BibitemShut {NoStop}%
\bibitem [{\citenamefont {Horodecki}\ and\ \citenamefont
  {Oppenheim}(2013)}]{horodecki2013fundamental}%
  \BibitemOpen
  \bibfield  {author} {\bibinfo {author} {\bibfnamefont {M.}~\bibnamefont
  {Horodecki}}\ and\ \bibinfo {author} {\bibfnamefont {J.}~\bibnamefont
  {Oppenheim}},\ }\href@noop {} {\bibfield  {journal} {\bibinfo  {journal}
  {Nature communications}\ }\textbf {\bibinfo {volume} {4}},\ \bibinfo {pages}
  {2059} (\bibinfo {year} {2013})}\BibitemShut {NoStop}%
\bibitem [{\citenamefont {Lostaglio}\ \emph {et~al.}(2015)\citenamefont
  {Lostaglio}, \citenamefont {Jennings},\ and\ \citenamefont
  {Rudolph}}]{LostaglioRudolphCohConstraint}%
  \BibitemOpen
  \bibfield  {author} {\bibinfo {author} {\bibfnamefont {M.}~\bibnamefont
  {Lostaglio}}, \bibinfo {author} {\bibfnamefont {D.}~\bibnamefont {Jennings}},
  \ and\ \bibinfo {author} {\bibfnamefont {T.}~\bibnamefont {Rudolph}},\
  }\href@noop {} {\bibfield  {journal} {\bibinfo  {journal} {Nature
  communications}\ }\textbf {\bibinfo {volume} {6}},\ \bibinfo {pages} {6383}
  (\bibinfo {year} {2015})}\BibitemShut {NoStop}%
\bibitem [{\citenamefont {Uzdin}(2017)}]{RU2017genCI}%
  \BibitemOpen
  \bibfield  {author} {\bibinfo {author} {\bibfnamefont {R.}~\bibnamefont
  {Uzdin}},\ }\href@noop {} {\bibfield  {journal} {\bibinfo  {journal}
  {Physical Review E}\ }\textbf {\bibinfo {volume} {96}},\ \bibinfo {pages}
  {032128} (\bibinfo {year} {2017})}\BibitemShut {NoStop}%
\bibitem [{\citenamefont {Peres}(2006)}]{PeresBook}%
  \BibitemOpen
  \bibfield  {author} {\bibinfo {author} {\bibfnamefont {A.}~\bibnamefont
  {Peres}},\ }\href@noop {} {\emph {\bibinfo {title} {Quantum theory: concepts
  and methods}}},\ Vol.~\bibinfo {volume} {57}\ (\bibinfo  {publisher}
  {Springer Science \& Business Media},\ \bibinfo {year} {2006})\BibitemShut
  {NoStop}%
\bibitem [{\citenamefont {Esposito}\ and\ \citenamefont {Van~den
  Broeck}(2011)}]{Esposito2011EPL2Law}%
  \BibitemOpen
  \bibfield  {author} {\bibinfo {author} {\bibfnamefont {M.}~\bibnamefont
  {Esposito}}\ and\ \bibinfo {author} {\bibfnamefont {C.}~\bibnamefont {Van~den
  Broeck}},\ }\href@noop {} {\bibfield  {journal} {\bibinfo  {journal} {EPL
  (Europhysics Letters)}\ }\textbf {\bibinfo {volume} {95}},\ \bibinfo {pages}
  {40004} (\bibinfo {year} {2011})}\BibitemShut {NoStop}%
\bibitem [{\citenamefont {Sagawa}(2012)}]{Sagawa2012second}%
  \BibitemOpen
  \bibfield  {author} {\bibinfo {author} {\bibfnamefont {T.}~\bibnamefont
  {Sagawa}},\ }\href@noop {} {\bibfield  {journal} {\bibinfo  {journal}
  {Lectures on Quantum Computing, Thermodynamics and Statistical Physics}\
  }\textbf {\bibinfo {volume} {8}},\ \bibinfo {pages} {127} (\bibinfo {year}
  {2012})}\BibitemShut {NoStop}%
\bibitem [{\citenamefont {Strasberg}\ \emph {et~al.}(2017)\citenamefont
  {Strasberg}, \citenamefont {Schaller}, \citenamefont {Brandes},\ and\
  \citenamefont {Esposito}}]{strasberg2017quantum}%
  \BibitemOpen
  \bibfield  {author} {\bibinfo {author} {\bibfnamefont {P.}~\bibnamefont
  {Strasberg}}, \bibinfo {author} {\bibfnamefont {G.}~\bibnamefont {Schaller}},
  \bibinfo {author} {\bibfnamefont {T.}~\bibnamefont {Brandes}}, \ and\
  \bibinfo {author} {\bibfnamefont {M.}~\bibnamefont {Esposito}},\ }\href@noop
  {} {\bibfield  {journal} {\bibinfo  {journal} {Physical Review X}\ }\textbf
  {\bibinfo {volume} {7}},\ \bibinfo {pages} {021003} (\bibinfo {year}
  {2017})}\BibitemShut {NoStop}%
\bibitem [{\citenamefont {Lostaglio}\ \emph {et~al.}(2017)\citenamefont
  {Lostaglio}, \citenamefont {Jennings},\ and\ \citenamefont
  {Rudolph}}]{lostaglio2017non-commutativity}%
  \BibitemOpen
  \bibfield  {author} {\bibinfo {author} {\bibfnamefont {M.}~\bibnamefont
  {Lostaglio}}, \bibinfo {author} {\bibfnamefont {D.}~\bibnamefont {Jennings}},
  \ and\ \bibinfo {author} {\bibfnamefont {T.}~\bibnamefont {Rudolph}},\
  }\href@noop {} {\bibfield  {journal} {\bibinfo  {journal} {New Journal of
  Physics}\ }\textbf {\bibinfo {volume} {19}},\ \bibinfo {pages} {043008}
  (\bibinfo {year} {2017})}\BibitemShut {NoStop}%
\bibitem [{\citenamefont {Guryanova}\ \emph {et~al.}(2016)\citenamefont
  {Guryanova}, \citenamefont {Popescu}, \citenamefont {Short}, \citenamefont
  {Silva},\ and\ \citenamefont {Skrzypczyk}}]{YelenaGGE}%
  \BibitemOpen
  \bibfield  {author} {\bibinfo {author} {\bibfnamefont {Y.}~\bibnamefont
  {Guryanova}}, \bibinfo {author} {\bibfnamefont {S.}~\bibnamefont {Popescu}},
  \bibinfo {author} {\bibfnamefont {A.~J.}\ \bibnamefont {Short}}, \bibinfo
  {author} {\bibfnamefont {R.}~\bibnamefont {Silva}}, \ and\ \bibinfo {author}
  {\bibfnamefont {P.}~\bibnamefont {Skrzypczyk}},\ }\href@noop {} {\bibfield
  {journal} {\bibinfo  {journal} {Nat Commun}\ }\textbf {\bibinfo {volume}
  {7}},\ \bibinfo {pages} {12049} (\bibinfo {year} {2016})}\BibitemShut
  {NoStop}%
\bibitem [{\citenamefont {Halpern}\ \emph {et~al.}(2016)\citenamefont
  {Halpern}, \citenamefont {Faist}, \citenamefont {Oppenheim},\ and\
  \citenamefont {Winter}}]{halpernGGErt}%
  \BibitemOpen
  \bibfield  {author} {\bibinfo {author} {\bibfnamefont {N.~Y.}\ \bibnamefont
  {Halpern}}, \bibinfo {author} {\bibfnamefont {P.}~\bibnamefont {Faist}},
  \bibinfo {author} {\bibfnamefont {J.}~\bibnamefont {Oppenheim}}, \ and\
  \bibinfo {author} {\bibfnamefont {A.}~\bibnamefont {Winter}},\ }\href@noop {}
  {\bibfield  {journal} {\bibinfo  {journal} {Nature Communications}\ }\textbf
  {\bibinfo {volume} {7}},\ \bibinfo {pages} {12051} (\bibinfo {year}
  {2016})}\BibitemShut {NoStop}%
\bibitem [{\citenamefont {Jarzynski}(1999)}]{JarzynskiMicroscopicClausius}%
  \BibitemOpen
  \bibfield  {author} {\bibinfo {author} {\bibfnamefont {C.}~\bibnamefont
  {Jarzynski}},\ }\href {\doibase 10.1023/A:1004541004050} {\bibfield
  {journal} {\bibinfo  {journal} {Journal of Statistical Physics}\ }\textbf
  {\bibinfo {volume} {96}},\ \bibinfo {pages} {415} (\bibinfo {year}
  {1999})}\BibitemShut {NoStop}%
\bibitem [{\citenamefont {Deffner}\ and\ \citenamefont
  {Jarzynski}(2013)}]{deffner2013information}%
  \BibitemOpen
  \bibfield  {author} {\bibinfo {author} {\bibfnamefont {S.}~\bibnamefont
  {Deffner}}\ and\ \bibinfo {author} {\bibfnamefont {C.}~\bibnamefont
  {Jarzynski}},\ }\href@noop {} {\bibfield  {journal} {\bibinfo  {journal}
  {Physical Review X}\ }\textbf {\bibinfo {volume} {3}},\ \bibinfo {pages}
  {041003} (\bibinfo {year} {2013})}\BibitemShut {NoStop}%
\bibitem [{\citenamefont {Deffner}\ and\ \citenamefont
  {Lutz}(2010)}]{DeffnerLutzRelEntBures}%
  \BibitemOpen
  \bibfield  {author} {\bibinfo {author} {\bibfnamefont {S.}~\bibnamefont
  {Deffner}}\ and\ \bibinfo {author} {\bibfnamefont {E.}~\bibnamefont {Lutz}},\
  }\href@noop {} {\bibfield  {journal} {\bibinfo  {journal} {Physical review
  letters}\ }\textbf {\bibinfo {volume} {105}},\ \bibinfo {pages} {170402}
  (\bibinfo {year} {2010})}\BibitemShut {NoStop}%
\bibitem [{\citenamefont {Nielsen}\ and\ \citenamefont
  {Chuang}(2002)}]{nielsen2002QuantInfo}%
  \BibitemOpen
  \bibfield  {author} {\bibinfo {author} {\bibfnamefont {M.~A.}\ \bibnamefont
  {Nielsen}}\ and\ \bibinfo {author} {\bibfnamefont {I.}~\bibnamefont
  {Chuang}},\ }\href@noop {} {\enquote {\bibinfo {title} {Quantum computation
  and quantum information},}\ } (\bibinfo {year} {2002})\BibitemShut {NoStop}%
\bibitem [{\citenamefont {Marshall}\ \emph {et~al.}(1979)\citenamefont
  {Marshall}, \citenamefont {Olkin},\ and\ \citenamefont
  {Arnold}}]{Marshall1979MajorizationBook}%
  \BibitemOpen
  \bibfield  {author} {\bibinfo {author} {\bibfnamefont {A.~W.}\ \bibnamefont
  {Marshall}}, \bibinfo {author} {\bibfnamefont {I.}~\bibnamefont {Olkin}}, \
  and\ \bibinfo {author} {\bibfnamefont {B.~C.}\ \bibnamefont {Arnold}},\
  }\href@noop {} {\emph {\bibinfo {title} {Inequalities: theory of majorization
  and its applications}}},\ Vol.\ \bibinfo {volume} {143}\ (\bibinfo
  {publisher} {Springer},\ \bibinfo {year} {1979})\BibitemShut {NoStop}%
\bibitem [{\citenamefont {{P. Oscar Boykin, Tal Mor, Vwani Roychowdhury,
  Farrokh Vatan, and Rutger Vrijen}}(2002)}]{tal02}%
  \BibitemOpen
  \bibfield  {author} {\bibinfo {author} {\bibnamefont {{P. Oscar Boykin, Tal
  Mor, Vwani Roychowdhury, Farrokh Vatan, and Rutger Vrijen}}},\ }\href@noop {}
  {\bibfield  {journal} {\bibinfo  {journal} {Proc. Nat. Acad. Sci.}\ }\textbf
  {\bibinfo {volume} {99}},\ \bibinfo {pages} {3388} (\bibinfo {year}
  {2002})}\BibitemShut {NoStop}%
\bibitem [{\citenamefont {Jarzynski}(2011)}]{Jarzynski2011equalitiesReview}%
  \BibitemOpen
  \bibfield  {author} {\bibinfo {author} {\bibfnamefont {C.}~\bibnamefont
  {Jarzynski}},\ }\href@noop {} {\bibfield  {journal} {\bibinfo  {journal}
  {Annu. Rev. Condens. Matter Phys.}\ }\textbf {\bibinfo {volume} {2}},\
  \bibinfo {pages} {329} (\bibinfo {year} {2011})}\BibitemShut {NoStop}%
\bibitem [{\citenamefont {Speck}\ and\ \citenamefont
  {Seifert}(2007)}]{Seifert2007FTRev}%
  \BibitemOpen
  \bibfield  {author} {\bibinfo {author} {\bibfnamefont {T.}~\bibnamefont
  {Speck}}\ and\ \bibinfo {author} {\bibfnamefont {U.}~\bibnamefont
  {Seifert}},\ }\href@noop {} {\bibfield  {journal} {\bibinfo  {journal}
  {Journal of Statistical Mechanics: Theory and Experiment}\ }\textbf {\bibinfo
  {volume} {2007}},\ \bibinfo {pages} {L09002} (\bibinfo {year}
  {2007})}\BibitemShut {NoStop}%
\bibitem [{\citenamefont {Harris}\ and\ \citenamefont
  {Sch{\"u}tz}(2007)}]{harris2007fluctuationReview}%
  \BibitemOpen
  \bibfield  {author} {\bibinfo {author} {\bibfnamefont {R.}~\bibnamefont
  {Harris}}\ and\ \bibinfo {author} {\bibfnamefont {G.}~\bibnamefont
  {Sch{\"u}tz}},\ }\href@noop {} {\bibfield  {journal} {\bibinfo  {journal}
  {Journal of Statistical Mechanics: Theory and Experiment}\ }\textbf {\bibinfo
  {volume} {2007}},\ \bibinfo {pages} {P07020} (\bibinfo {year}
  {2007})}\BibitemShut {NoStop}%
\bibitem [{\citenamefont {Campisi}\ \emph {et~al.}(2015)\citenamefont
  {Campisi}, \citenamefont {Pekola},\ and\ \citenamefont
  {Fazio}}]{campisi2014FT_SolidStateExp}%
  \BibitemOpen
  \bibfield  {author} {\bibinfo {author} {\bibfnamefont {M.}~\bibnamefont
  {Campisi}}, \bibinfo {author} {\bibfnamefont {J.}~\bibnamefont {Pekola}}, \
  and\ \bibinfo {author} {\bibfnamefont {R.}~\bibnamefont {Fazio}},\
  }\href@noop {} {\bibfield  {journal} {\bibinfo  {journal} {New Journal of
  Physics}\ }\textbf {\bibinfo {volume} {17}},\ \bibinfo {pages} {035012}
  (\bibinfo {year} {2015})}\BibitemShut {NoStop}%
\bibitem [{\citenamefont {Jarzynski}(1997)}]{jarzynski1997nonequilibrium}%
  \BibitemOpen
  \bibfield  {author} {\bibinfo {author} {\bibfnamefont {C.}~\bibnamefont
  {Jarzynski}},\ }\href@noop {} {\bibfield  {journal} {\bibinfo  {journal}
  {Physical Review Letters}\ }\textbf {\bibinfo {volume} {78}},\ \bibinfo
  {pages} {2690} (\bibinfo {year} {1997})}\BibitemShut {NoStop}%
\bibitem [{\citenamefont {Micadei}\ \emph {et~al.}(2017)\citenamefont
  {Micadei}, \citenamefont {Peterson}, \citenamefont {Souza}, \citenamefont
  {Sarthour}, \citenamefont {Oliveira}, \citenamefont {Landi}, \citenamefont
  {Batalh{\~a}o}, \citenamefont {Serra},\ and\ \citenamefont
  {Lutz}}]{SerraLutz2spinNMR}%
  \BibitemOpen
  \bibfield  {author} {\bibinfo {author} {\bibfnamefont {K.}~\bibnamefont
  {Micadei}}, \bibinfo {author} {\bibfnamefont {J.~P.}\ \bibnamefont
  {Peterson}}, \bibinfo {author} {\bibfnamefont {A.~M.}\ \bibnamefont {Souza}},
  \bibinfo {author} {\bibfnamefont {R.~S.}\ \bibnamefont {Sarthour}}, \bibinfo
  {author} {\bibfnamefont {I.~S.}\ \bibnamefont {Oliveira}}, \bibinfo {author}
  {\bibfnamefont {G.~T.}\ \bibnamefont {Landi}}, \bibinfo {author}
  {\bibfnamefont {T.~B.}\ \bibnamefont {Batalh{\~a}o}}, \bibinfo {author}
  {\bibfnamefont {R.~M.}\ \bibnamefont {Serra}}, \ and\ \bibinfo {author}
  {\bibfnamefont {E.}~\bibnamefont {Lutz}},\ }\href@noop {} {\bibfield
  {journal} {\bibinfo  {journal} {arXiv preprint arXiv:1711.03323}\ } (\bibinfo
  {year} {2017})}\BibitemShut {NoStop}%
\bibitem [{\citenamefont {Ro{\ss}nagel}\ \emph {et~al.}(2014)\citenamefont
  {Ro{\ss}nagel}, \citenamefont {Abah}, \citenamefont {Schmidt-Kaler},
  \citenamefont {Singer},\ and\ \citenamefont {Lutz}}]{LutzSqueezedBaths}%
  \BibitemOpen
  \bibfield  {author} {\bibinfo {author} {\bibfnamefont {J.}~\bibnamefont
  {Ro{\ss}nagel}}, \bibinfo {author} {\bibfnamefont {O.}~\bibnamefont {Abah}},
  \bibinfo {author} {\bibfnamefont {F.}~\bibnamefont {Schmidt-Kaler}}, \bibinfo
  {author} {\bibfnamefont {K.}~\bibnamefont {Singer}}, \ and\ \bibinfo {author}
  {\bibfnamefont {E.}~\bibnamefont {Lutz}},\ }\href@noop {} {\bibfield
  {journal} {\bibinfo  {journal} {Physical review letters}\ }\textbf {\bibinfo
  {volume} {112}},\ \bibinfo {pages} {030602} (\bibinfo {year}
  {2014})}\BibitemShut {NoStop}%
\bibitem [{\citenamefont {Manzano}\ \emph {et~al.}(2015)\citenamefont
  {Manzano}, \citenamefont {Galve}, \citenamefont {Zambrini},\ and\
  \citenamefont {Parrondo}}]{ParrondoSqueezedBath}%
  \BibitemOpen
  \bibfield  {author} {\bibinfo {author} {\bibfnamefont {G.}~\bibnamefont
  {Manzano}}, \bibinfo {author} {\bibfnamefont {F.}~\bibnamefont {Galve}},
  \bibinfo {author} {\bibfnamefont {R.}~\bibnamefont {Zambrini}}, \ and\
  \bibinfo {author} {\bibfnamefont {J.~M.}\ \bibnamefont {Parrondo}},\
  }\href@noop {} {\bibfield  {journal} {\bibinfo  {journal} {arXiv preprint
  arXiv:1512.07881}\ } (\bibinfo {year} {2015})}\BibitemShut {NoStop}%
\bibitem [{\citenamefont {Maslennikov}\ \emph {et~al.}(2017)\citenamefont
  {Maslennikov}, \citenamefont {Ding}, \citenamefont {Hablutzel}, \citenamefont
  {Gan}, \citenamefont {Roulet}, \citenamefont {Nimmrichter}, \citenamefont
  {Dai}, \citenamefont {Scarani},\ and\ \citenamefont
  {Matsukevich}}]{Roulet_3_Ion_fridge2017}%
  \BibitemOpen
  \bibfield  {author} {\bibinfo {author} {\bibfnamefont {G.}~\bibnamefont
  {Maslennikov}}, \bibinfo {author} {\bibfnamefont {S.}~\bibnamefont {Ding}},
  \bibinfo {author} {\bibfnamefont {R.}~\bibnamefont {Hablutzel}}, \bibinfo
  {author} {\bibfnamefont {J.}~\bibnamefont {Gan}}, \bibinfo {author}
  {\bibfnamefont {A.}~\bibnamefont {Roulet}}, \bibinfo {author} {\bibfnamefont
  {S.}~\bibnamefont {Nimmrichter}}, \bibinfo {author} {\bibfnamefont
  {J.}~\bibnamefont {Dai}}, \bibinfo {author} {\bibfnamefont {V.}~\bibnamefont
  {Scarani}}, \ and\ \bibinfo {author} {\bibfnamefont {D.}~\bibnamefont
  {Matsukevich}},\ }\href@noop {} {\bibfield  {journal} {\bibinfo  {journal}
  {arXiv preprint arXiv:1702.08672}\ } (\bibinfo {year} {2017})}\BibitemShut
  {NoStop}%
\bibitem [{\citenamefont {Niedenzu}\ \emph {et~al.}(2018)\citenamefont
  {Niedenzu}, \citenamefont {Mukherjee}, \citenamefont {Ghosh}, \citenamefont
  {Kofman},\ and\ \citenamefont {Kurizki}}]{Wolgang2018passivityCI}%
  \BibitemOpen
  \bibfield  {author} {\bibinfo {author} {\bibfnamefont {W.}~\bibnamefont
  {Niedenzu}}, \bibinfo {author} {\bibfnamefont {V.}~\bibnamefont {Mukherjee}},
  \bibinfo {author} {\bibfnamefont {A.}~\bibnamefont {Ghosh}}, \bibinfo
  {author} {\bibfnamefont {A.~G.}\ \bibnamefont {Kofman}}, \ and\ \bibinfo
  {author} {\bibfnamefont {G.}~\bibnamefont {Kurizki}},\ }\href@noop {}
  {\bibfield  {journal} {\bibinfo  {journal} {Nature communications}\ }\textbf
  {\bibinfo {volume} {9}},\ \bibinfo {pages} {165} (\bibinfo {year}
  {2018})}\BibitemShut {NoStop}%
\bibitem [{\citenamefont {{A. E. Allahverdyan, R. Balian, and Th. M.
  Nieuwenhuizen}}(2004)}]{AllahverdyanErgotropy}%
  \BibitemOpen
  \bibfield  {author} {\bibinfo {author} {\bibnamefont {{A. E. Allahverdyan, R.
  Balian, and Th. M. Nieuwenhuizen}}},\ }\href@noop {} {\bibfield  {journal}
  {\bibinfo  {journal} {Euro. Phys. Lett.}\ }\textbf {\bibinfo {volume} {67}},\
  \bibinfo {pages} {{565}} (\bibinfo {year} {2004})}\BibitemShut {NoStop}%
\bibitem [{\citenamefont {Perarnau-Llobet}\ \emph {et~al.}(2017)\citenamefont
  {Perarnau-Llobet}, \citenamefont {B{\"a}umer}, \citenamefont {Hovhannisyan},
  \citenamefont {Huber},\ and\ \citenamefont {Acin}}]{Marti2017NoGoQWorkFluch}%
  \BibitemOpen
  \bibfield  {author} {\bibinfo {author} {\bibfnamefont {M.}~\bibnamefont
  {Perarnau-Llobet}}, \bibinfo {author} {\bibfnamefont {E.}~\bibnamefont
  {B{\"a}umer}}, \bibinfo {author} {\bibfnamefont {K.~V.}\ \bibnamefont
  {Hovhannisyan}}, \bibinfo {author} {\bibfnamefont {M.}~\bibnamefont {Huber}},
  \ and\ \bibinfo {author} {\bibfnamefont {A.}~\bibnamefont {Acin}},\
  }\href@noop {} {\bibfield  {journal} {\bibinfo  {journal} {Physical review
  letters}\ }\textbf {\bibinfo {volume} {118}},\ \bibinfo {pages} {070601}
  (\bibinfo {year} {2017})}\BibitemShut {NoStop}%
\bibitem [{\citenamefont {Bera}\ \emph {et~al.}(2017)\citenamefont {Bera},
  \citenamefont {Riera}, \citenamefont {Lewenstein},\ and\ \citenamefont
  {Winter}}]{BeraWinters2ndLawCat}%
  \BibitemOpen
  \bibfield  {author} {\bibinfo {author} {\bibfnamefont {M.~N.}\ \bibnamefont
  {Bera}}, \bibinfo {author} {\bibfnamefont {A.}~\bibnamefont {Riera}},
  \bibinfo {author} {\bibfnamefont {M.}~\bibnamefont {Lewenstein}}, \ and\
  \bibinfo {author} {\bibfnamefont {A.}~\bibnamefont {Winter}},\ }\href@noop {}
  {\bibfield  {journal} {\bibinfo  {journal} {arXiv preprint arXiv:1707.01750}\
  } (\bibinfo {year} {2017})}\BibitemShut {NoStop}%
\bibitem [{\citenamefont {Streltsov}\ \emph {et~al.}(2017)\citenamefont
  {Streltsov}, \citenamefont {Adesso},\ and\ \citenamefont
  {Plenio}}]{PlenioAdesso2017coherenceRev}%
  \BibitemOpen
  \bibfield  {author} {\bibinfo {author} {\bibfnamefont {A.}~\bibnamefont
  {Streltsov}}, \bibinfo {author} {\bibfnamefont {G.}~\bibnamefont {Adesso}}, \
  and\ \bibinfo {author} {\bibfnamefont {M.~B.}\ \bibnamefont {Plenio}},\
  }\href@noop {} {\bibfield  {journal} {\bibinfo  {journal} {Reviews of Modern
  Physics}\ }\textbf {\bibinfo {volume} {89}},\ \bibinfo {pages} {041003}
  (\bibinfo {year} {2017})}\BibitemShut {NoStop}%
\bibitem [{\citenamefont {Uzdin}\ and\ \citenamefont
  {Rahav}(2018)}]{GlobalPassivity}%
  \BibitemOpen
  \bibfield  {author} {\bibinfo {author} {\bibfnamefont {R.}~\bibnamefont
  {Uzdin}}\ and\ \bibinfo {author} {\bibfnamefont {S.}~\bibnamefont {Rahav}},\
  }\href@noop {} {\bibfield  {journal} {\bibinfo  {journal} {Phys. Rev. X, In
  press, arXiv 1805.00220}\ } (\bibinfo {year} {2018})}\BibitemShut {NoStop}%
\bibitem [{\citenamefont {Kammerlander}\ and\ \citenamefont
  {Anders}(2016)}]{Anders2015MeasurementWork}%
  \BibitemOpen
  \bibfield  {author} {\bibinfo {author} {\bibfnamefont {P.}~\bibnamefont
  {Kammerlander}}\ and\ \bibinfo {author} {\bibfnamefont {J.}~\bibnamefont
  {Anders}},\ }\href@noop {} {\bibfield  {journal} {\bibinfo  {journal}
  {Scientific Reports}\ }\textbf {\bibinfo {volume} {6}},\ \bibinfo {pages}
  {22174} (\bibinfo {year} {2016})}\BibitemShut {NoStop}%
\bibitem [{\citenamefont {Still}\ \emph {et~al.}(2012)\citenamefont {Still},
  \citenamefont {Sivak}, \citenamefont {Bell},\ and\ \citenamefont
  {Crooks}}]{CrooksThemoPredNonEf}%
  \BibitemOpen
  \bibfield  {author} {\bibinfo {author} {\bibfnamefont {S.}~\bibnamefont
  {Still}}, \bibinfo {author} {\bibfnamefont {D.~A.}\ \bibnamefont {Sivak}},
  \bibinfo {author} {\bibfnamefont {A.~J.}\ \bibnamefont {Bell}}, \ and\
  \bibinfo {author} {\bibfnamefont {G.~E.}\ \bibnamefont {Crooks}},\
  }\href@noop {} {\bibfield  {journal} {\bibinfo  {journal} {Physical review
  letters}\ }\textbf {\bibinfo {volume} {109}},\ \bibinfo {pages} {120604}
  (\bibinfo {year} {2012})}\BibitemShut {NoStop}%
\bibitem [{\citenamefont {Pusz}\ and\ \citenamefont {Wornwicz}(1978)}]{pusz78}%
  \BibitemOpen
  \bibfield  {author} {\bibinfo {author} {\bibfnamefont {W.}~\bibnamefont
  {Pusz}}\ and\ \bibinfo {author} {\bibfnamefont {S.}~\bibnamefont
  {Wornwicz}},\ }\href@noop {} {\bibfield  {journal} {\bibinfo  {journal}
  {Commun. Math. Phys.}\ }\textbf {\bibinfo {volume} {58}},\ \bibinfo {pages}
  {273} (\bibinfo {year} {1978})}\BibitemShut {NoStop}%
\bibitem [{\citenamefont {Lenard}(1978)}]{lenard1978Gibbs}%
  \BibitemOpen
  \bibfield  {author} {\bibinfo {author} {\bibfnamefont {A.}~\bibnamefont
  {Lenard}},\ }\href@noop {} {\bibfield  {journal} {\bibinfo  {journal}
  {Journal of Statistical Physics}\ }\textbf {\bibinfo {volume} {19}},\
  \bibinfo {pages} {575} (\bibinfo {year} {1978})}\BibitemShut {NoStop}%
\bibitem [{\citenamefont {Perarnau-Llobet}\ \emph
  {et~al.}(2015{\natexlab{a}})\citenamefont {Perarnau-Llobet}, \citenamefont
  {Hovhannisyan}, \citenamefont {Huber}, \citenamefont {Skrzypczyk},
  \citenamefont {Tura},\ and\ \citenamefont
  {Ac{\'\i}n}}]{Marti2015EnergeticPassive}%
  \BibitemOpen
  \bibfield  {author} {\bibinfo {author} {\bibfnamefont {M.}~\bibnamefont
  {Perarnau-Llobet}}, \bibinfo {author} {\bibfnamefont {K.~V.}\ \bibnamefont
  {Hovhannisyan}}, \bibinfo {author} {\bibfnamefont {M.}~\bibnamefont {Huber}},
  \bibinfo {author} {\bibfnamefont {P.}~\bibnamefont {Skrzypczyk}}, \bibinfo
  {author} {\bibfnamefont {J.}~\bibnamefont {Tura}}, \ and\ \bibinfo {author}
  {\bibfnamefont {A.}~\bibnamefont {Ac{\'\i}n}},\ }\href@noop {} {\bibfield
  {journal} {\bibinfo  {journal} {Physical Review E}\ }\textbf {\bibinfo
  {volume} {92}},\ \bibinfo {pages} {042147} (\bibinfo {year}
  {2015}{\natexlab{a}})}\BibitemShut {NoStop}%
\bibitem [{\citenamefont {Perarnau-Llobet}\ \emph
  {et~al.}(2015{\natexlab{b}})\citenamefont {Perarnau-Llobet}, \citenamefont
  {Hovhannisyan}, \citenamefont {Huber}, \citenamefont {Skrzypczyk},
  \citenamefont {Brunner},\ and\ \citenamefont {Ac\'{\i}n}}]{MartiWorkCorr}%
  \BibitemOpen
  \bibfield  {author} {\bibinfo {author} {\bibfnamefont {M.}~\bibnamefont
  {Perarnau-Llobet}}, \bibinfo {author} {\bibfnamefont {K.~V.}\ \bibnamefont
  {Hovhannisyan}}, \bibinfo {author} {\bibfnamefont {M.}~\bibnamefont {Huber}},
  \bibinfo {author} {\bibfnamefont {P.}~\bibnamefont {Skrzypczyk}}, \bibinfo
  {author} {\bibfnamefont {N.}~\bibnamefont {Brunner}}, \ and\ \bibinfo
  {author} {\bibfnamefont {A.}~\bibnamefont {Ac\'{\i}n}},\ }\href@noop {}
  {\bibfield  {journal} {\bibinfo  {journal} {Phys. Rev. X}\ }\textbf {\bibinfo
  {volume} {5}},\ \bibinfo {pages} {041011} (\bibinfo {year}
  {2015}{\natexlab{b}})}\BibitemShut {NoStop}%
\bibitem [{\citenamefont {{Amikam Levy and Ronnie Kosloff}}(2012)}]{k272}%
  \BibitemOpen
  \bibfield  {author} {\bibinfo {author} {\bibnamefont {{Amikam Levy and Ronnie
  Kosloff}}},\ }\href@noop {} {\bibfield  {journal} {\bibinfo  {journal} {Phys.
  Rev. Lett.}\ }\textbf {\bibinfo {volume} {108}},\ \bibinfo {pages} {070604}
  (\bibinfo {year} {2012})}\BibitemShut {NoStop}%
\bibitem [{\citenamefont {Kirkwood}(1935)}]{kirkwood1935PMF}%
  \BibitemOpen
  \bibfield  {author} {\bibinfo {author} {\bibfnamefont {J.~G.}\ \bibnamefont
  {Kirkwood}},\ }\href@noop {} {\bibfield  {journal} {\bibinfo  {journal} {The
  Journal of Chemical Physics}\ }\textbf {\bibinfo {volume} {3}},\ \bibinfo
  {pages} {300} (\bibinfo {year} {1935})}\BibitemShut {NoStop}%
\bibitem [{\citenamefont
  {Jarzynski}(2017)}]{Jarzynski2017_PRX_strong_coupling}%
  \BibitemOpen
  \bibfield  {author} {\bibinfo {author} {\bibfnamefont {C.}~\bibnamefont
  {Jarzynski}},\ }\href@noop {} {\bibfield  {journal} {\bibinfo  {journal}
  {Physical Review X}\ }\textbf {\bibinfo {volume} {7}},\ \bibinfo {pages}
  {011008} (\bibinfo {year} {2017})}\BibitemShut {NoStop}%
\bibitem [{\citenamefont {Seifert}(2016)}]{SeifertPRL2016_strong_coupling}%
  \BibitemOpen
  \bibfield  {author} {\bibinfo {author} {\bibfnamefont {U.}~\bibnamefont
  {Seifert}},\ }\href@noop {} {\bibfield  {journal} {\bibinfo  {journal}
  {Physical review letters}\ }\textbf {\bibinfo {volume} {116}},\ \bibinfo
  {pages} {020601} (\bibinfo {year} {2016})}\BibitemShut {NoStop}%
\bibitem [{\citenamefont {Miller}\ and\ \citenamefont
  {Anders}(2017)}]{JanetPRE2017_strong_coupling}%
  \BibitemOpen
  \bibfield  {author} {\bibinfo {author} {\bibfnamefont {H.~J.~D.}\
  \bibnamefont {Miller}}\ and\ \bibinfo {author} {\bibfnamefont
  {J.}~\bibnamefont {Anders}},\ }\href {\doibase 10.1103/PhysRevE.95.062123}
  {\bibfield  {journal} {\bibinfo  {journal} {Phys. Rev. E}\ }\textbf {\bibinfo
  {volume} {95}},\ \bibinfo {pages} {062123} (\bibinfo {year}
  {2017})}\BibitemShut {NoStop}%
\bibitem [{\citenamefont {Anders}\ and\ \citenamefont
  {Giovannetti}(2013)}]{anders2013thermodynamics}%
  \BibitemOpen
  \bibfield  {author} {\bibinfo {author} {\bibfnamefont {J.}~\bibnamefont
  {Anders}}\ and\ \bibinfo {author} {\bibfnamefont {V.}~\bibnamefont
  {Giovannetti}},\ }\href@noop {} {\bibfield  {journal} {\bibinfo  {journal}
  {New Journal of Physics}\ }\textbf {\bibinfo {volume} {15}},\ \bibinfo
  {pages} {033022} (\bibinfo {year} {2013})}\BibitemShut {NoStop}%
\bibitem [{\citenamefont {Uzdin}()}]{ruXmachines}%
  \BibitemOpen
  \bibfield  {author} {\bibinfo {author} {\bibfnamefont {R.}~\bibnamefont
  {Uzdin}},\ }\href@noop {} {\bibinfo  {journal} {In preperation}\
  }\BibitemShut {NoStop}%
\bibitem [{\citenamefont {Brask}\ \emph {et~al.}(2015)\citenamefont {Brask},
  \citenamefont {Haack}, \citenamefont {Brunner},\ and\ \citenamefont
  {Huber}}]{HuberEntangGenHeat}%
  \BibitemOpen
\bibfield  {journal} {  }\bibfield  {author} {\bibinfo {author} {\bibfnamefont
  {J.~B.}\ \bibnamefont {Brask}}, \bibinfo {author} {\bibfnamefont
  {G.}~\bibnamefont {Haack}}, \bibinfo {author} {\bibfnamefont
  {N.}~\bibnamefont {Brunner}}, \ and\ \bibinfo {author} {\bibfnamefont
  {M.}~\bibnamefont {Huber}},\ }\href@noop {} {\bibfield  {journal} {\bibinfo
  {journal} {New Journal of Physics}\ }\textbf {\bibinfo {volume} {17}},\
  \bibinfo {pages} {113029} (\bibinfo {year} {2015})}\BibitemShut {NoStop}%
\bibitem [{\citenamefont {Tacchino}\ \emph {et~al.}(2018)\citenamefont
  {Tacchino}, \citenamefont {Auff{\`e}ves}, \citenamefont {Santos},\ and\
  \citenamefont {Gerace}}]{Alexia2018steady}%
  \BibitemOpen
  \bibfield  {author} {\bibinfo {author} {\bibfnamefont {F.}~\bibnamefont
  {Tacchino}}, \bibinfo {author} {\bibfnamefont {A.}~\bibnamefont
  {Auff{\`e}ves}}, \bibinfo {author} {\bibfnamefont {M.}~\bibnamefont
  {Santos}}, \ and\ \bibinfo {author} {\bibfnamefont {D.}~\bibnamefont
  {Gerace}},\ }\href@noop {} {\bibfield  {journal} {\bibinfo  {journal}
  {Physical review letters}\ }\textbf {\bibinfo {volume} {120}},\ \bibinfo
  {pages} {063604} (\bibinfo {year} {2018})}\BibitemShut {NoStop}%
\bibitem [{\citenamefont {Leff}\ and\ \citenamefont
  {Rex}(2014)}]{leff2014MaxwellDemonBook}%
  \BibitemOpen
  \bibfield  {author} {\bibinfo {author} {\bibfnamefont {H.~S.}\ \bibnamefont
  {Leff}}\ and\ \bibinfo {author} {\bibfnamefont {A.~F.}\ \bibnamefont {Rex}},\
  }\href@noop {} {\emph {\bibinfo {title} {Maxwell's demon: entropy,
  information, computing}}}\ (\bibinfo  {publisher} {Princeton University
  Press},\ \bibinfo {year} {2014})\BibitemShut {NoStop}%
\bibitem [{\citenamefont {Goold}\ \emph {et~al.}(2015)\citenamefont {Goold},
  \citenamefont {Paternostro},\ and\ \citenamefont
  {Modi}}]{Goold2015nonEq_Landuer}%
  \BibitemOpen
  \bibfield  {author} {\bibinfo {author} {\bibfnamefont {J.}~\bibnamefont
  {Goold}}, \bibinfo {author} {\bibfnamefont {M.}~\bibnamefont {Paternostro}},
  \ and\ \bibinfo {author} {\bibfnamefont {K.}~\bibnamefont {Modi}},\
  }\href@noop {} {\bibfield  {journal} {\bibinfo  {journal} {Physical review
  letters}\ }\textbf {\bibinfo {volume} {114}},\ \bibinfo {pages} {060602}
  (\bibinfo {year} {2015})}\BibitemShut {NoStop}%
\bibitem [{\citenamefont {Santos}\ \emph {et~al.}(2017)\citenamefont {Santos},
  \citenamefont {Landi},\ and\ \citenamefont
  {Paternostro}}]{Paternostro2017EntProdWigner}%
  \BibitemOpen
  \bibfield  {author} {\bibinfo {author} {\bibfnamefont {J.~P.}\ \bibnamefont
  {Santos}}, \bibinfo {author} {\bibfnamefont {G.~T.}\ \bibnamefont {Landi}}, \
  and\ \bibinfo {author} {\bibfnamefont {M.}~\bibnamefont {Paternostro}},\
  }\href@noop {} {\bibfield  {journal} {\bibinfo  {journal} {Physical review
  letters}\ }\textbf {\bibinfo {volume} {118}},\ \bibinfo {pages} {220601}
  (\bibinfo {year} {2017})}\BibitemShut {NoStop}%
\bibitem [{\citenamefont {Sparaciari}\ \emph {et~al.}(2017)\citenamefont
  {Sparaciari}, \citenamefont {Jennings},\ and\ \citenamefont
  {Oppenheim}}]{Carlo2017passive2thermal}%
  \BibitemOpen
  \bibfield  {author} {\bibinfo {author} {\bibfnamefont {C.}~\bibnamefont
  {Sparaciari}}, \bibinfo {author} {\bibfnamefont {D.}~\bibnamefont
  {Jennings}}, \ and\ \bibinfo {author} {\bibfnamefont {J.}~\bibnamefont
  {Oppenheim}},\ }\href@noop {} {\bibfield  {journal} {\bibinfo  {journal}
  {Nature communications}\ }\textbf {\bibinfo {volume} {8}},\ \bibinfo {pages}
  {1895} (\bibinfo {year} {2017})}\BibitemShut {NoStop}%
\end{thebibliography}%

\end{document}